\documentclass[
    aip,
    pop,
    amsmath,
    amssymb,
    reprint,
    floatfix
]{revtex4-1}
\usepackage{graphicx}
\usepackage{dcolumn}
\usepackage[mathlines]{lineno}

\usepackage[utf8]{inputenc}
\usepackage[T1]{fontenc}
\usepackage{mathptmx}
\usepackage{etoolbox}

\draft{}

\usepackage{amssymb}
\usepackage{amsfonts}
\usepackage{graphicx}
\usepackage{multirow}
\usepackage{bigstrut}
\usepackage{epstopdf}
\usepackage{upgreek}
\usepackage[english]{babel}
\usepackage{csquotes}
\usepackage{combelow}
\usepackage{amsmath}
\usepackage{amsthm}
\usepackage{xcolor}
\usepackage{bm}
\usepackage{tikz}
\usetikzlibrary{shapes}
\usetikzlibrary{arrows}
\usetikzlibrary{positioning}
\usetikzlibrary{matrix}
\usetikzlibrary{patterns}
\usetikzlibrary{decorations.pathreplacing,decorations.pathmorphing,decorations.markings}
\usepackage{tikz-qtree}
\usetikzlibrary{trees,calc,arrows.meta,positioning,bending}
\usepackage{mathtools}
\usepackage{hyperref}

\newcommand{\R}{\mathbb{R}}

\newcommand{\argmin}{\mathop{\mathrm{argmin}}\limits}

\hypersetup{
    pdftitle={Scientific Machine Learning Based Reduced-Order Models for Plasma Turbulence Simulations},
    pdfauthor={Constantin Gahr, Ionut-Gabriel Farcas, Frank Jenko},
    pdfkeywords={Hasegawa-Wakatani equations, data-driven reduced modeling, scientific machine learning, reduced models, reduced models Hasegawa-Wakatani equations, data-driven reduced models Hasegawa-Wakatani equations}
}

\newcommand{\Vr}{\mathbf{V}_r}

\newcommand{\yb}{\mathbf{y}}
\newcommand{\Yb}{\mathbf{Y}}

\newcommand{\qb}{\mathbf{q}}
\newcommand{\Qb}{\mathbf{Q}}

\newcommand{\Ah}{\hat{\mathbf{A}}}

\newcommand{\Hh}{\hat{\mathbf{H}}}
\newcommand{\Gh}{\hat{\mathbf{G}}}

\newcommand{\qh}{\hat{\mathbf{q}}}
\newcommand{\yh}{\hat{\mathbf{y}}}

\begin{document}

\title{Scientific Machine Learning Based Reduced-Order Models for Plasma Turbulence Simulations}

\author{Constantin Gahr}
\email{constantin.gahr@ipp.mpg.de}
\affiliation{Max Planck Institute for Plasma Physics, 85748 Garching, Germany}
\altaffiliation{C.~Gahr and I.~Farca\c{s} contributed equally to this work}

\author{Ionu\c{t}-Gabriel Farca\c{s}}
\email{farcasi@vt.edu}
\affiliation{Oden Institute for Computational Engineering and Sciences, The University of Texas at Austin, Austin, Texas 78712}
\affiliation{Department of Mathematics, Virginia Tech, Blacksburg, Virginia 24061}
\altaffiliation{C.~Gahr and I.~Farca\c{s} contributed equally to this work}

\author{Frank Jenko}
\email{frank.jenko@ipp.mpg.de}
\affiliation{Max Planck Institute for Plasma Physics, 85748 Garching, Germany}

\begin{abstract}
    This paper investigates non-intrusive Scientific Machine Learning (SciML) Reduced-Order Models (ROMs) for plasma turbulence simulations.
    In particular, we focus on Operator Inference (OpInf) to build low-cost physics-based ROMs from data for such simulations.
    As a representative example, we consider the (classical) Hasegawa-Wakatani (HW) equations used for modeling two-dimensional electrostatic drift-wave turbulence.
    For a comprehensive perspective of the potential of OpInf to construct predictive ROMs, we consider three setups for the HW equations by varying a key parameter, namely the adiabaticity coefficient.
    These setups lead to the formation of complex and nonlinear dynamics, which makes the construction of predictive ROMs of any kind challenging.
    We generate the training datasets by performing direct numerical simulations of the HW equations and recording the computed state data and outputs the over a time horizon of $100$ time units in the turbulent phase.
    We then use these datasets to construct OpInf ROMs for predictions over $400$ additional time units, that is, $400\%$ more than the training horizon.
    Our results show that the OpInf ROMs capture important statistical features of the turbulent dynamics and generalize beyond the training time horizon while reducing the computational effort of the high-fidelity simulation by up to five orders of magnitude.
    In the broader context of fusion research, this shows that non-intrusive SciML ROMs have the potential to drastically accelerate numerical studies, which can ultimately enable tasks such as the design of optimized fusion devices.
\end{abstract}


\maketitle 

\section{Introduction}\label{sec:intro}

Understanding, predicting, and controlling turbulent transport in magnetic confinement devices is critical toward commercially viable fusion.
High-fidelity simulations of turbulent transport, despite tremendous progress, remain impractical for many design or control tasks due to their high computational cost even on powerful supercomputers.
This necessitates the development of innovative reduced modeling approaches to enable these tasks at scale.

Simulations of plasma turbulence in fusion devices have been revolutionized through high-performance computing, facilitating the transition from merely qualitative to quantitative and detailed predictions.
Furthermore, the recent emergence of exascale-capable machines~\cite{atchleyFrontierExploringExascale2023} paves the way toward full-device~\cite{germaschewskiExascaleWholedeviceModeling2021} and complex multiscale turbulent transport simulations~\cite{belliSpectralTransitionMultiscale2022}.
However, these simulations often require significant resources even on large supercomputers, thus restricting the number of cases that can be studied explicitly.
This, in turn, prohibits the routine use of high-fidelity models for many-query tasks that require ensembles of simulations such as the design and control of optimized fusion devices, or uncertainty quantification and sensitivity analysis~\cite{farcasGeneralFrameworkQuantifying2022}.
An alternative is provided by Reduced-Order Models (ROMs)~\cite{bennerModelReductionApproximation2017, bennerSurveyProjectionBasedModel2015} which aim to construct computationally cheap but sufficiently accurate approximations of high-fidelity models with the goal of replacing them for the aforementioned tasks.

In recent years, Scientific Machine Learning (SciML) developed tools for building ROMs from data for complex applications by combining the rigor of physics-based modeling with the convenience of data-driven learning.
In essence, given data stemming from numerical simulations or experimental measurements (or both), SciML constructs ROMs by embedding physics principles into the learning problem.

The present paper focuses on constructing SciML ROMs for complex, nonlinear plasma turbulence models.
Our goal is to show that SciML ROMs can provide statistically accurate predictions that capture the important features of the underlying problem while reducing the computational cost of the high-fidelity simulation code by orders of magnitude.
In particular, we consider Operator Inference (OpInf)~\cite{peherstorferDatadrivenOperatorInference2016, kramerLearningNonlinearReduced2024}.
OpInf is a SciML approach for learning physics-based reduced models from data for models with polynomial structure.
More general types of nonlinearities can also be considered by using lifting transformations~\cite{kramerNonlinearModelOrder2019, qianLiftLearnPhysicsinformed2020} that expose polynomial structure in the lifted governing equations.
The potential of OpInf for constructing accurate ROMs was demonstrated across a wide range of applications, including complex reactive flows in rocket combustion~\cite{farcasDistributedComputingPhysicsbased2024, farcasParametricNonintrusiveReducedorder2023, qianReducedOperatorInference2022, swischukLearningPhysicsbasedReducedorder2020}, solar wind predictions in space weather applications~\cite{issanPredictingSolarWind2023}, and solidification simulations~\cite{khodabakhshiNonintrusiveDatadrivenModel2022}.
Additionally, Ref.~\cite{almeidaNonIntrusiveReducedModels2022} demonstrated promising forecasting capabilities for the state of canonical chaotic systems such as Lorenz 96 and the Kuramoto-Sivashinsky equations.
Extensions include using filtering~\cite{farcasFilteringNonintrusiveDatadriven2022} and roll-outs~\cite{uyOperatorInferenceRoll2023} to handle noisy, scarce, and low-quality data, localization~\cite{geelenLocalizedNonintrusiveReducedorder2022}, domain decomposition~\cite{farcasImprovingAccuracyScalability2023}, and quadratic manifolds~\cite{geelenOperatorInferenceNonintrusive2023} to address some of the challenges in reducing problems with complex dynamics.
The recent review paper~\cite{kramerLearningNonlinearReduced2024} provides a comprehensive overview on OpInf.

The plasma turbulence model under consideration is given by the (classical) Hasegawa-Wakatani (HW) equations~\cite{hasegawaPlasmaEdgeTurbulence1983}.
They model two-dimensional electrostatic drift-wave turbulence in a slab geometry and have been studied extensively~\cite{camargoResistiveDriftWave1995, manzMicroscopicPicturePlasma2019, numataBifurcationElectrostaticResistive2007}.
The dynamics of these equations were analyzed using the roper orthogonal decomposition (POD)~\cite{sasakiEvaluationAbruptEnergy2020, yatomiDatadrivenModalAnalysis2023}, a classical ROM approach~\cite{berkoozProperOrthogonalDecomposition1993}.
Furthermore, Ref.~\cite{goumiriReducedorderModelBased2013} constructed ROMs for the linearized HW equations using balanced truncation.
Recently, there has been a surge in research dedicated to developing neural network approximations for the HW equations, including initial turbulent state predictions~\cite{castagnaStyleGANAIDeconvolution2024}, closure models for Large Eddy Simulations~\cite{greifPhysicsPreservingAIAcceleratedSimulations2023}, surrogate models for transport simulations~\cite{clavierGenerativeMachineLearning2024}, direct particle flux predictions~\cite{heinonenTurbulenceModelReduction2020}, and synthesized impurity structures~\cite{linSynthesizingImpurityClustering2024}.
Further studies related to plasma physics beyond the HW equations include Refs.~\cite{farajiDynamicModeDecomposition2023, futataniSpatiotemporalMultiscalingAnalysis2009, kusabaSparsityPromotingDynamicMode2020, taylorDynamicModeDecomposition2018} which used dynamic mode decomposition (DMD) and POD for analyzing plasma dynamics, Refs.~\cite{farajiDynamicModeDecomposition2023II, kaptanogluCharacterizingMagnetizedPlasmas2020} which used these methods for constructing ROMs for predictions beyond their training time horizon, Ref.~\cite{jajimaEstimation2DProfile2023} which employed neural networks to approximate quantities of interest such as the particle flux, and Ref.~\cite{gopakumarFourierNeuralOperator2023} which used neural operators to construct surrogates.
Beyond plasma turbulence, Ref.~\cite{ghazijahaniPredictionTurbulentFlow2024} used a POD-based Echo State Network to approximate 2D turbulent transport modeled via the Navier-Stokes equation using a setup similar to ours.
Nonetheless, constructing accurate and predictive ROMs for complex problems exhibiting chaotic behavior such as the HW equations remains an open challenge.

We take one step further in this direction and investigate the potential of OpInf for constructing computationally cheap SciML ROMs for forecasting the state and key quantities of interest for the HW equations, namely the particle flux $\Gamma_n$ and the resistive dissipation rate $\Gamma_c$.
These two quantities of interest represent the dominant energy source and sink in the HW model and are therefore ubiquitous for assessing turbulent transport.
To mimic realistic scenarios where training datasets are often expensive to obtain, we consider a setup where training data are limited to a short time horizon, while the ROM is used for predictions over a longer time horizon.
Additionally, to gain a comprehensive understanding of OpInf in this context, we explore three distinct turbulent regimes by varying the adiabaticity coefficient $c_1$, which influences the coupling between fields in the HW model.
Given the nonlinear and chaotic nature of the HW equations, constructing ROMs that provide point-wise accurate predictions (for both the state and quantities of interest) is extremely challenging with most standard model reduction methods.
Recognizing that statistically accurate predictions are often sufficient in practical applications, our primary goal is to demonstrate that OpInf can provide stable predictions over long time horizons that accurately capture the statistical properties of the two quantities of interest.
To further evaluate the particle transport in the state predictions from the OpInf ROM,
we also compute the phase shift between the predicted states.
Our findings reveal that the OpInf ROMs can indeed generate stable and statistically accurate predictions of the quantities of interest, as well as state predictions that accurately approximate the low wavenumbers of the phase shift.
We note that to the best of our knowledge, this work represents one of the first studies in which SciML ROMs are considered for the HW equations.

The remainder of the paper is organized as follows.
Section~\ref{sec:hw-overview} summarizes the HW equations, the relevant quantities of interest, and the setup for high-fidelity numerical simulations.
Section~\ref{sec:contributions} presents the steps to construct physics-based data-driven ROMs via OpInf for the HW system.
Section~\ref{sec:results} presents our numerical results and discusses our findings.
We conclude the paper in Sec.~\ref{sec:conclusions}.

\section{Modeling plasma turbulence via the Hasegawa-Wakatani equations}\label{sec:hw-overview}

This section summarizes the (classical) HW equations for modeling two-dimensional electrostatic drift-wave turbulence.
Section~\ref{subsec:hw-background} provides the physics background.
Section~\ref{subsec:hw-equation} presents the HW equations, followed by Sec.~\ref{subsec:hw-invariants}, which presents the two key quantities of interest and defines the phase shift.
Finally, Sec.~\ref{subsec:hw_training_data} provides the setup used for solving the HW equations numerically.

\subsection{Physics background}\label{subsec:hw-background}

We consider the (classical) HW equations, which represent a nonlinear two-dimensional fluid model for drift-wave turbulence in magnetized plasmas~\cite{camargoResistiveDriftWave1995, hasegawaPlasmaEdgeTurbulence1983, wakataniCollisionalDriftWave1984}.
These equations model the time evolution of plasma density and potential fluctuations in a 2D slab geometry of size $L \times L$ under the influence of the background magnetic field and background density gradient.

In this model, the magnetic field is assumed to be homogeneous and constant in the $z$-direction.
The background density $n_0(x, y) = n_0(x)$ is non-uniform in the $x$-direction and has a constant equilibrium density scale $L_n = | \partial_x \log n_0 |^{-1} $.
The gradient in the background density transports particles in the positive $x$-direction, which continuously introduces energy and drives the turbulent transport in the system.

The spatial and temporal coordinates and the fields are rescaled to be dimensionless.
Following Ref.~\cite{camargoResistiveDriftWave1995}, let $\bar{x}$, $\bar{y}$, and $\bar{t}$ represent the two spatial and time coordinates in Gaussian Centimeter-Gram-Second (CGS) units.
The inverse of the sound radius $\rho_s = c_s / \omega_{ci}$ scales the spatial coordinates to $x = \bar{x} /\rho_s$ and $y = \bar{y} / \rho_s$, where $c_s = \sqrt{T_e / m_i}$ denotes the ion sound speed and $\omega_{ci} = e B_0 / (m_i c)$ is the ion cyclotron frequency.
$T_e$ denotes the electron temperature, $m_i$ represents the ion mass, $e$ is the electric charge, $B_0$ is the magnetic field strength, and $c$ is the speed of light.
The electron drift frequency $\omega_{de} = c_s / L_n$ scales the time component to $t = \bar{t}\omega_{de}$.
The normalized density and potential fluctuations $\tilde{n}$ and $\tilde{\phi}$ are obtained from the original fields $\bar{n}$ and $\bar{\phi}$ as
\begin{equation}
    \tilde{n} = \frac{\bar{n}}{n_0} \frac{L_n}{\rho_s} \quad \text{and} \quad \tilde{\phi} = \frac{e \bar{\phi}}{T_e} \frac{L_n}{\rho_s}.
\end{equation}

\subsection{The Hasegawa-Wakatani equations}\label{subsec:hw-equation}

The (classical) HW equations read
\begin{equation}\label{eq:hw}
    \begin{cases}
        {\partial_t} \nabla^2 \tilde{\phi} & = c_1(\tilde{\phi} - \tilde{n}) - \{\tilde{\phi}, \nabla^2 \tilde{\phi} \} + \nu \nabla^{2N} \nabla^{2}\tilde{\phi}    \\
        {\partial_t} \tilde{n}             & = c_1 (\tilde{\phi} - \tilde{n}) - \{ \tilde{\phi}, \tilde{n}\} - \partial_y \tilde{\phi} + \nu \nabla^{2N} \tilde{n},
    \end{cases}
\end{equation}
where $\nabla^2 \tilde{\phi}$ is the vorticity and $\{f, g\} := \partial_x f \partial_y g - \partial_y f \partial_x g$ denotes the Poisson bracket of $f$ and $g$.
The adiabaticity parameter $c_1 > 0$ controls how fast the electric potential reacts to changes in the density and exerts a significant impact on the dynamics of these equations.
In the limit $c_1 \to \infty$, the potential reacts instantaneously to changes in the density and the model reduces to the Hasegawa-Mima equation~\cite{hasegawaPseudoThreeDimensional1978}.
The other extreme, $c_1 \to 0$, decouples the density and potential fluctuations, reducing the model to the Navier-Stokes equations~\cite{camargoResistiveDriftWave1995}.
The hyperdiffusion terms $\nabla^{2N}$ are introduced to prevent energy accumulation at the grid scale, with hyperdiffusion coefficient $\nu$ that controls the amount of dissipated energy.
Periodic boundary conditions are assumed in both spatial directions.
After initialization, the dynamics undergo a transient phase in which linear drift-waves dominate and the energy grows exponentially.
The dynamics subsequently enter the turbulent phase, where the effect of quadratic nonlinearities becomes significant and turbulent transport dominates.

\subsection{Energy balance, quantities of interest and phase shift}\label{subsec:hw-invariants}

The energy $E$ of the HW equations is defined as
\begin{equation}
    E(t) = \frac{1}{2} \int \! \mathrm{d} \mathbf{x} \; \; \left( \tilde{n}{(t, \mathbf{x})}^2 + | \nabla \tilde{\phi}(t, \mathbf{x}) |^2 \right),
\end{equation}
where $\mathbf{x} = (x, y)$ and $\mathrm{d}\mathbf{x} = \mathrm{d}x\,\mathrm{d}y$\cite{camargoResistiveDriftWave1995}.
Its time evolution reads
\begin{equation}
    \frac{\partial E}{\partial t} =  \Gamma_n - \Gamma_c - \mathcal{D}^E,
\end{equation}
where
\begin{subequations}\label{eq:invariants}
    \begin{align}
        \Gamma_n(t)      & = - \int \! \mathrm{d}\mathbf{x} \; \;  \tilde{n}(t, \mathbf{x}) \partial_y \tilde{\phi}(t, \mathbf{x}), \label{eq:gamma_n_def}                                                                            \\
        \Gamma_c(t)      & = c_1 \int \! \mathrm{d}\mathbf{x} \; {\left( \tilde{n}(t, \mathbf{x}) - \tilde{\phi}(t, \mathbf{x}) \right)}^2, \label{eq:gamma_c_def}                                                                    \\
        \mathcal{D}^E(t) & = \nu \int \! \mathrm{d} \mathbf{x} \; \; \left( \tilde{n}(t, \mathbf{x}) \nabla^{2N} \tilde{n}(t, \mathbf{x}) - \tilde{\phi}(t, \mathbf{x})\nabla^{2N + 2} \tilde{\phi}(t, \mathbf{x}) \right) \nonumber.
    \end{align}
\end{subequations}
Equation~\eqref{eq:gamma_n_def} defines the particle flux $\Gamma_n$, which measures the rate at which particles are transported from the background density $n_0$ to the fluctuating density $\tilde{n}$.
It constitutes the only energy source in the HW equations and drives the turbulent dynamics.
At the same time, energy is resistively dissipated, which is quantified by the
resistive dissipation rate $\Gamma_c$ defined in Eq.~\eqref{eq:gamma_c_def}.
$\Gamma_c$ and the viscous dissipation rate $\mathcal{D}^E$ represent the only energy sinks in the HW equations.
The latter, however, is generally small.
For $c_1 \geq 1$, we have $\mathcal{D}^E / \Gamma_n < 1 \% $\cite{camargoResistiveDriftWave1995} and $\Gamma_n \sim \Gamma_c$, and $\mathcal{D}^E$ becomes negligible as $c_1 \to \infty$.

The particle flux $\Gamma_n$ and the resistive dissipation rate $\Gamma_c$ therefore represent the dominant energy source and sink in the HW model.
$\Gamma_n$ and $\Gamma_c$ are the two quantities of interest in our ROM experiments.
Additionally, for a more in-depth assessment of our ROMs and, in particular, of the particle transport in the ROM forecasted states, we also consider the phase shift $\delta_\mathbf{k}(t)$ between density and potential fluctuations~\cite{camargoResistiveDriftWave1995}.
The phase shift represents a key measure of the particle transport~\cite{kimNumericalInvestigationFrequency2013} and is related to the adiabatic response of the fluctuations.
It is defined as the complex angle
\begin{equation}\label{eq:phase_shift}
    \delta_\mathbf{k}(t) = \mathrm{Im} \, \log \left( \tilde{n}_\mathbf{k}^\star{(t)} \, \tilde{\phi}_\mathbf{k}(t) \right) ,
\end{equation}
where $\tilde{n}_\mathbf{k}(t) = \mathcal{F}{[\tilde{n}]}_\mathbf{k}(t)$ and $\tilde{\phi}_\mathbf{k}(t) = \mathcal{F}{[\tilde{\phi}]}_\mathbf{k}(t)$ denote the Fourier transforms of density and potential fluctuations, respectively, $\mathbf{k} = (k_x, k_y)$ is the wavenumber, and $\cdot^\star$ denotes the complex conjugate.
Similarly to Ref.~\cite{kimNumericalInvestigationFrequency2013}, we compute the temporal mean of the phase shift~\eqref{eq:phase_shift} at $k_x = 0$ to quantify how well the state predictions obtained with our ROMs capture the underlying particle transport and adiabatic response.

\subsection{Setup for Direct Numerical Simulations}\label{subsec:hw_training_data}
We solve the HW equations~\eqref{eq:hw} numerically using a Direct Numerical Simulation (DNS)~\cite{greifHW2DReferenceImplementation2023}.
The computational domain has a $2$D slab geometry of size $L \times L = {(2\pi/k_0)}^2$ with $k_0 = 0.15$.
It is discretized using an equidistant grid comprising $N_x = 512 \times 512 = 262,144$ spatial degrees of freedom.
Since we have $N_s = 2$ state variables, the total number of degrees of freedom is $N_{\mathrm{state}} = N_s \times N_x = 524,288$.

The Poisson brackets are discretized using a fourth--order Arakawa scheme~\cite{arakawaComputationalDesignLongTerm1997}.
The remaining spatial derivatives, including the one in Eq.~\eqref{eq:gamma_n_def}, are discretized using second--order finite differences.
The time iterations consist of two steps: first, the semi-discrete HW equations~\eqref{eq:hw} are solved using the explicit $4$th-order Runge-Kutta method~\cite{rungeUeberNumerischeAufloesung1895} to compute the density~$\tilde{n}$ and vorticity~$\nabla^2 \tilde{\phi}$, followed by solving the Poisson equation~$\nabla^2 \tilde{\phi} \to \tilde{\phi}$ via spectral methods to compute~$\tilde{\phi}$.
The $N_{\mathrm{out}} = 2$ quantities of interest~\eqref{eq:gamma_n_def} and~\eqref{eq:gamma_c_def} are then estimated using the trapezoidal rule while the phase shift~\eqref{eq:phase_shift} is computed using the discrete Fourier transform.
The equations are solved in increments of $\delta_t = 0.01$.
The initial conditions are instances of a 2D Gaussian random field with zero mean and covariance $0.01 \times \mathrm{Id}$.
The diffusivity parameter is $\nu = 5 \cdot 10^{-9}$ with hyperdiffusion order $N = 3$.
A reference implementation is available at~\url{https://github.com/the-rccg/hw2d}~\cite{greifHW2DReferenceImplementation2023}.

\begin{figure*}[t]
    \centering
    \includegraphics{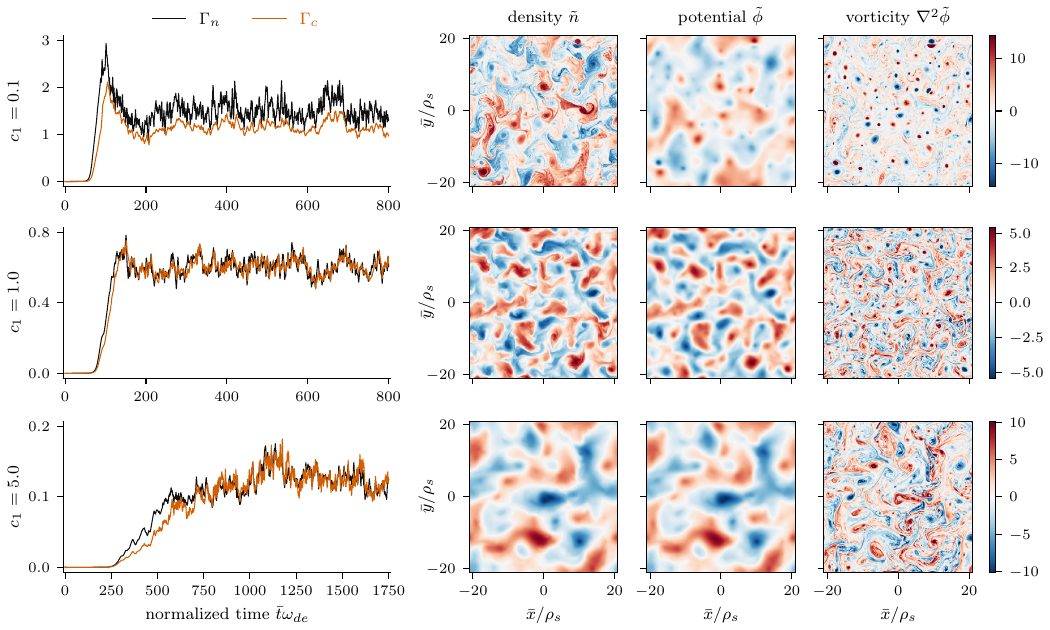}
    \caption{\label{fig:reference_invariants_snapshots}
        Particle flux $\Gamma_n$ and resistive dissipation rate $\Gamma_c$ from initialization to nonlinear saturation (left), and snapshots of density fluctuations~$\tilde{n}$, potential fluctuations~$\tilde{\phi}$, and normalized vorticity~$\nabla^2 \tilde{\phi}$ (right).
    }
\end{figure*}

To obtain a comprehensive perspective, we consider three values for the adiabaticity parameter $c_1$, namely $c_1=0.1$ (close to the hydrodynamic limit), the typically studied intermediate value $c_1 = 1.0$, and $c_1 = 5.0$ (with adiabatic electrons toward the Hasegawa-Mima limit).
Figure~\ref{fig:reference_invariants_snapshots}, on the right, shows examples of density, potential, and (normalized) vorticity fields in the turbulent regime.
Note that the similarity between density and potential increases with $c_1$.
All $c_1$ values lead to the formation of complex, nonlinear and self-driven dynamics, rendering the construction of accurate and predictive ROMs of any kind extremely challenging.

The three DNS were performed over a sufficiently long time horizon to ensure that we have enough simulation data within the turbulent phase for a thorough assessment of the predictive capabilities of the ROMs.
Figure~\ref{fig:reference_invariants_snapshots}, on the left, plots the two quantities of interest starting from $t = 0$ for the three values of $c_1$.
The onset of turbulent-driven transport increases with $c_1$.
Because of this, the DNS for $c_1 = 0.1$ and $c_1 = 1.0$ were performed over time horizon $[0, 800]$, each requiring about $16$ hours on a single core of an Intel(R) Xeon(R) 6148 CPU.\@
The DNS for $c_1 = 5.0$ was performed over $[0; 1,750]$, necessitating roughly $35$ hours on a single core.
In our OpInf experiments, we will focus solely on the turbulent phase since this is generally the primary region of interest, and disregard the initial and transient stages.
In particular, we consider time horizons where the turbulent dynamics are fully developed, which are $t \geq 300$ for $c_1 = 0.1$ and $c_1 = 1.0$, and $t \geq 1,250$ for $c_1 = 5.0$.

\section{Learning physics-based reduced models from data for the Hasegawa-Wakatani system via Operator Inference}\label{sec:contributions}
We propose using OpInf to construct computationally inexpensive data-driven ROMs for the HW system.
Section~\ref{subsec:opinf_for_hw_preliminaries} summarizes the OpInf setup, followed by Sec.~\ref{subsec:opinf_for_hw_steps} which presents the steps to construct OpInf ROMs.

\subsection{Preliminaries}\label{subsec:opinf_for_hw_preliminaries}
We start by identifying the structure of the governing equations.
In OpInf, the governing equations are considered in semi-discrete form after discretizing the underlying PDE model in space.
Using the steps summarized in Sec.~\ref{subsec:hw_training_data}, the spatial discretization of the HW equations~\eqref{eq:hw} and the two quantities of interest given by Eqs.~\eqref{eq:gamma_n_def} and~\eqref{eq:gamma_c_def}, is a linear-quadratic system of coupled equations,
\begin{subequations}\label{eq:FOM_quad_model}
    \begin{align}
        \dot{\qb}(t) & = \mathbf{A}\qb(t) + \mathbf{H} (\qb(t) \otimes \qb(t)) \label{eq:FOM_quad_model_state} \\
        \yb(t)       & = \mathbf{G} (\qb(t) \otimes \qb(t)),
    \end{align}
\end{subequations}
where $\qb(t) = \begin{pmatrix} \tilde{\mathbf{n}}(t) & \tilde{\mathbf{\phi}}(t) \end{pmatrix}^\top \in \mathbb{R}^{N_{\mathrm{state}}}$ with $N_{\mathrm{state}} = 524,288$ denotes the spatially discretized state variables at time $t$, and $\yb(t) = \begin{pmatrix} \Gamma_n(t) & \Gamma_c(t)  \end{pmatrix}^\top \in \mathbb{R}^{N_{\mathrm{out}}}$ with $N_{\mathrm{out}} = 2$ the corresponding outputs.
The quadratic operator $\mathbf{H} \in \mathbb{R}^{N_{\mathrm{state}} \times N_{\mathrm{state}}^2}$ stems from the discretized Poisson brackets in Eq.~\eqref{eq:hw}, while the linear operator $\mathbf{A} \in \mathbb{R}^{N_{\mathrm{state}} \times N_{\mathrm{state}}}$ originates from the remaining, linear terms.
The quantities of interest defined in Eqs.~\eqref{eq:gamma_n_def} and~\eqref{eq:gamma_c_def} depend quadratically on the state via the quadratic operator $\mathbf{G} \in \mathbb{R}^{N_{\mathrm{out}} \times N_{\mathrm{state}}^2}$.
We refer to Eq.~\eqref{eq:FOM_quad_model} as the \emph{HW system}.
We note that OpInf does not require to explicitly form the linear-quadratic system of equations~\eqref{eq:FOM_quad_model}; OpInf is a non-intrusive, data-driven approach that only necessitates knowledge about the structure of the governing equations.

We focus on the time evolution of the HW system over a time horizon $[t_\text{init}, t_\text{final}]$ in the fully developed turbulent phase, where $t_\text{init} = 300$ for $c_1 = 0.1$ and $c_1 = 1.0$, and $t_\text{init} = 1,250$ for $c_1 = 5.0$.
Moreover, $t_\text{final}$ denotes the final time.
The initial condition in~\eqref{eq:FOM_quad_model_state} is the high-dimensional state solution at $t = t_\text{init}$.
We consider the setup where we are given a training dataset over a time horizon $[t_\text{init}, t_\text{train}]$ with $t_\text{train} < t_\text{final}$ obtained by solving the HW system using the setup from Sec.~\ref{subsec:hw_training_data} and recording the corresponding state vectors $\qb_i = \qb(t_i)$ and outputs $\yb_i = \yb(t_i)$ at $n_t$ time instants $t_i \in [t_\text{init}, t_\text{train}]$.
The states and outputs are assembled into a snapshot matrix $\Qb \in \mathbb{R}^{N_{\mathrm{state}} \times n_t}$ and an output matrix $\Yb \in \mathbb{R}^{N_{\mathrm{out}} \times n_t}$ as
\begin{equation*}
    \Qb =
    \begin{bmatrix}
        \vert & \vert &        & \vert     \\
        \qb_1 & \qb_2 & \cdots & \qb_{n_t} \\
        \vert & \vert &        & \vert
    \end{bmatrix}
    \quad \text{and} \quad
    \Yb =
    \begin{bmatrix}
        \vert & \vert &        & \vert     \\
        \yb_1 & \yb_2 & \cdots & \yb_{n_t} \\
        \vert & \vert &        & \vert
    \end{bmatrix},
\end{equation*}
with the state vectors $\qb_i$ and outputs $\yb_i$ as their respective $i$th columns.
Given $\Qb$ and $\Yb$, our goal is to construct  computationally inexpensive OpInf ROMs that capture the statistical properties of the outputs beyond the training time horizon and provide state predictions that approximate well the phase shift.

\subsection{Learning physics-based data-driven reduced models via Operator Inference}\label{subsec:opinf_for_hw_steps}
OpInf is a SciML approach that combines the perspectives of physics-based modeling with data-driven learning to non-intrusively construct computationally cheap ROMs of systems with polynomial nonlinearities from data~\cite{peherstorferDatadrivenOperatorInference2016}.
For more general types of nonlinearities, lifting transformations~\cite{qianLiftLearnPhysicsinformed2020} can be used to expose polynomial structure in the lifted governing equations.
OpInf requires two main ingredients: knowledge about the structure of the governing equations and a training data set to construct a structure-preserving ROM.\@
The HW system~\eqref{eq:FOM_quad_model} is quadratic in both state and output, therefore not necessitating lifting transformations.
The steps to construct quadratic ROMs via OpInf are outlined below.

The snapshot matrix $\Qb$ is used to compute a rank-$r$ linear POD basis to represent the high-dimensional state snapshots in the low-dimensional subspace spanned by the POD basis vectors.
One of the most prevalent methods for determining this basis uses the thin singular value decomposition of the snapshot matrix~\cite{golubMatrixComputations2013}
\begin{equation*}
    \Qb = \mathbf{V} \bm{\Sigma} \mathbf{W}^\top,
\end{equation*}
where $\mathbf{V} \in \R^{N_{\mathrm{state}} \times n_t}$ contains the left singular vectors and $\mathbf{W} \in \R^{n_t \times n_t}$ the right singular vectors.
$\bm{\Sigma} \in \R^{n_t \times n_t}$ is a diagonal matrix with the singular values arranged on the diagonal in non-increasing order, that is, $\sigma_i \geq \sigma_{i+1}$ for $i = 1, 2, \ldots, n_t - 1$.
The rank-$r$ POD basis $\Vr \in \mathbb{R}^{N_{\mathrm{state}} \times r}$ consists of the first $r \ll N_{\mathrm{state}}$ columns of $\mathbf{V}$, i.e., the left singular vectors corresponding to the $r$ largest singular values.
We note that in problems with multiple state variables with different scales, it is recommended to center and scale the training snapshots variable-by-variable prior to computing the basis.
In case of the HW equations, density $\tilde{n}$ and potential $\tilde{\phi}$ have similar scales and data transformations are not necessary.

The representation of the high-dimensional state snapshots in the low-dimensional subspace spanned by the POD basis vectors is computed as
\begin{equation}\label{eq:opinf_projection}
    \qh_i = \Vr^\top \qb_i \in \R^r
\end{equation}
for $i = 1, 2, \ldots, n_t$.
The reduced states $\qh_i$ are then recorded as columns into a matrix $\hat{\mathbf{Q}} \in \mathbb{R}^{r \times n_t}$.
The dimension of the output $\Yb$ remains unchanged.
Analogously to the state data, multidimensional outputs with different scales may need to be centered and scaled prior to constructing ROMs.

OpInf then learns the reduced operators for a structure-preserving quadratic ROM,
\begin{subequations}\label{eq:opinf_quad_model}
    \begin{align}
        \dot{\qh}(t) & = \Ah \qh(t) + \Hh (\qh(t) \otimes \qh(t)) \label{eq:opinf_quad_model_state} \\
        \yh(t)       & = \Gh (\qh(t) \otimes \qh(t)), \label{eq:opinf_quad_model_output}
    \end{align}
\end{subequations}
where $\Ah \in \mathbb{R}^{r \times r}$, $\Hh \in \mathbb{R}^{r \times r^2}$, and $\Gh \in \mathbb{R}^{N_{\mathrm{out}} \times r^2}$, and $\yh(t) \in \mathbb{R}^{N_{\mathrm{out}}}$ denotes the output ROM approximation.
The reduced operators are inferred by solving two least-square minimizations, one for the reduced state dynamics
\begin{equation}\label{eq:rom_min_problem_state}
    \argmin_{\Ah, \hat{\mathbf{H}}}  \sum_{i = 1}^{n_t} \left\lVert \dot{\hat{\qb}}_i - \Ah \qh_i  - \hat{\mathbf{H}} (\qh_i \otimes \qh_i) \right\rVert_2^2 + \beta_A \| \Ah \|_F^2 + \beta_H \| \hat{\mathbf{H}} \|_F^2,
\end{equation}
and the other for the outputs
\begin{equation}\label{eq:rom_min_problem_output}
    \argmin_{\Gh}  \sum_{i = 1}^{n_t} \left\lVert \yh_i - \Gh (\qh_i \otimes \qh_i)  \right\rVert_2^2 + \beta_G \| \Gh \|_F^2,
\end{equation}
where $\| \cdot \|_2$ and $\| \cdot \|_F$ denote the $2$- and Frobenius norms, respectively, while $\beta_A > 0$, $\beta_H > 0$, and $\beta_G > 0$ are regularization parameters.
Moreover, $\dot{\qh}_i \in \mathbb{R}^r$ denotes the projected time derivative of the $i$th snapshot vector.
If the numerical code provides the high-dimensional time derivative $\dot{\qb}_i \in \mathbb{R}^{N}$, $\dot{\qh}_i$ is obtained as $\dot{\qh}_i = \mathbf{V}_r^\top \dot{\qb}_i$.
Otherwise, it must be estimated numerically via finite differences, for example.
However, least-squares minimizations are known to be sensitive to right-hand-side perturbations~\cite{golubMatrixComputations2013}, hence an inaccurate approximation of $\dot{\qh}_i$ will have a deleterious influence on solution, especially in chaotic systems such as the HW equations considered in the present work.
In such cases, using the time-discrete formulation of OpInf is preferred~\cite{farcasFilteringNonintrusiveDatadriven2022}.
Therein, the derivative $\dot{\qh}_i$ in Eq.~\eqref{eq:rom_min_problem_state} is replaced by the time-shifted state $\qh_{i + 1}$, similarly to DMD.\@

Following Ref.~\cite{mcquarrieDatadrivenReducedorderModels2021}, Tikhonov regularization with separate regularization hyperparameters for linear and quadratic reduced operators is used to reduce overfitting or to compensate for other possible errors and model misspecifications.
It plays a key role in constructing predictive OpInf ROMs in problems with complex dynamics, and it will be crucial in the present work as well.
Refs.~\cite{mcquarrieDatadrivenReducedorderModels2021, qianReducedOperatorInference2022} proposed determining the regularization hyperparameters in Eq.~\eqref{eq:rom_min_problem_state} via a two-dimensional grid search over candidate values for each hyperparameter.
The optimal regularization pair is the one that minimizes the training error subject to the constraint that the learned reduced coefficients have bounded growth within a trial time horizon $[t_\text{init}, t_\text{trial}]$ with $t_\text{trial} \geq t_\text{final}$.
This can be extended to also finding the optimal regularization hyperparameter for the reduced output equation~\eqref{eq:rom_min_problem_output}.
However, deterministic procedures that aim to find a single regularization set of parameter may struggle with chaotic systems such as the HW system.

We propose an alternative, statistical strategy that computes an ensemble of solutions.
We consider a set of candidate regularization pairs $\mathcal{B}_{\mathrm{state}} \subset \mathbb{R}_{> 0}^2$ for~\eqref{eq:rom_min_problem_state}, and a set of candidate regularization parameters $\mathcal{B}_{\mathrm{out}} \subset \mathbb{R}_{> 0}$ for~\eqref{eq:rom_min_problem_output}.
For each element of $(\beta_A, \beta_H) \in \mathcal{B}_{\mathrm{state}}$, we solve~\eqref{eq:rom_min_problem_state} to find the reduced state operators, which are used to compute the reduced state solution over the target time horizon $[t_\text{init}, t_\text{final}]$ by solving~\eqref{eq:opinf_quad_model_state}.
Let $\tilde{\mathbf{Q}} \in \mathbb{R}^{r \times (n_t + n_p)}$ comprise the computed reduced state solutions in its columns, where $n_t + n_p$ denotes the total number of time instants over $[t_\text{init}, t_\text{final}]$.
We then compute the reduced output operator by solving~\eqref{eq:rom_min_problem_output} for each candidate $\beta_G \in \mathcal{B}_{\mathrm{out}}$, and plug the reduced state solutions $\tilde{\mathbf{Q}}$ into the reduced output equation~\eqref{eq:opinf_quad_model_output} to approximate the outputs over $[t_\text{init}, t_\text{final}]$.
Note that these steps are computationally cheap as they depend on the reduced dimension $r \ll N_\text{state}$.
Next, we extract the two quantities of interest from the approximate outputs and compute their respective means and standard deviations over both training and prediction time horizons.
As noted previously, the approximate quantities of interest are expected to differ point-wisely from the corresponding high-fidelity reference solutions over the prediction horizon due to the nonlinear and chaotic nature of the HW system.
Exploiting the fact that in the turbulent phase the statistics of the quantities of interest over the training and prediction horizons are similar, we compute the relative errors of their means and standard deviations with respect to the means and standard deviations of the reference solutions over the training horizon.
The ensemble comprises all approximate outputs with relative errors falling below user-prescribed tolerances.
We can then compute the ensemble mean as well as the standard deviation to assess the variability of the ensemble solutions within the considered regularization bounds.

Finally, given an approximate reduced state solution $\tilde{\mathbf{Q}}$ computed via OpInf over the target time horizon $[t_\text{init}, t_\text{final}]$, the phase shift defined in Eq.~\eqref{eq:phase_shift} can be estimated as follows: (i) map $\tilde{\mathbf{Q}}$ to the original coordinates by computing $\mathbf{V}_r \tilde{\mathbf{Q}}$, where $\mathbf{V}_r$ denotes the rank-$r$ POD basis, (ii) extract the components of the density and potential from the mapped solution, and (iii) compute their Fourier transforms and plug them into Eq.~\eqref{eq:phase_shift}.

Before moving forward, we note that recent works in the context of fluid turbulence (see, for example,~\cite{liMultiscaleReconstructionTurbulent2023, ghazijahaniPredictionTurbulentFlow2024} and the references therein) argued that using least-square error losses, similarly to OpInf, can result in surrogate or reduced models that might fail to capture complex features in turbulent and other complex, multi-scale scenarios.
This motivates the development of methods that use alternative loss functions such as the Kullback-Leibler divergence or the Wasserstein-Fourier metric~\cite{cazellesWassersteinFourierDistanceStationary2021} as well as nonlinear methods that go beyond linear POD bases~\cite{brunaNeuralGalerkinSchemes2024,peherstorferModelReductionTransportDominated2020,gopakumarFourierNeuralOperator2023}.
We leave such investigations for our future research.
As stated previously, our goal in this work is to investigate the potential of OpInf for constructing computationally cheap SciML ROMs for forecasting the state and key quantities of interest for the HW equations.

\section{Numerical experiments and discussion}\label{sec:results}

We employ the described OpInf procedure to learn physics-based ROMs for the HW system.
In Sec.~\ref{subsec:results_opinf_setup} we summarize the setup for OpInf while Section~\ref{subsec:opinf_results} reports our results.
In Sec.~\ref{subsec:opinf_sensitivity}, we discuss an important aspect related to constructing ROMs for chaotic systems, namely the sensitivity of the ROM predictions to changes in the initial condition used to generate the training data.
All ROM calculations were performed on a shared-memory machine with $256$ AMD EPYC 7702 CPUs and $2$ TB of RAM, using the \texttt{numpy}~\cite{harrisArrayProgrammingNumPy2020} and \texttt{scipy}~\cite{virtanenSciPyFundamentalAlgorithms2020} scientific computing libraries in \texttt{python}.

\subsection{Setup for Operator Inference}\label{subsec:results_opinf_setup}

We construct OpInf ROMs for $c_1 \in \{0.1, 1.0, 5.0\}$.
The training datasets are obtained by solving the governing equations and computing the corresponding outputs using the setup summarized in Sec.~\ref{subsec:hw_training_data}, and recording the computed high-dimensional state vectors and outputs at every second time iteration in the turbulent regime (i.e., we downsample the simulation results by a factor of two).
The initial conditions for the three DNS are distinct realization of the mean-free 2D Gaussian random field with covariance $0.01 \times \mathrm{Id}$.
Recall that $t_{\mathrm{init}} = 300$ for $c_1 = 0.1$ and $c_1 = 1.0$, and $t_{\mathrm{init}} = 1,250$ for $c_1 = 5.0$.
We consider the setup in which training data are provided for $100$ time units, for which we have $n_t = 5,000$ state snapshots and outputs.
Therefore, $t_{\mathrm{train}} = 400$ for $c_1 = 0.1$ and $c_1 = 1.0$, and $t_{\mathrm{train}} = 1,350$ for $c_1 = 5.0$,
Moreover, the corresponding snapshot and output matrices have dimensions $524,288 \times 5,000$ and $2 \times 5,000$, respectively.
We focus on using limited training data to simulate practical constraints in acquiring training datasets.
Our experiments with smaller training sets revealed that reduced data availability compromised the predictive capabilities of the resulting ROMs.

\begin{figure*}
    \centering
    \includegraphics{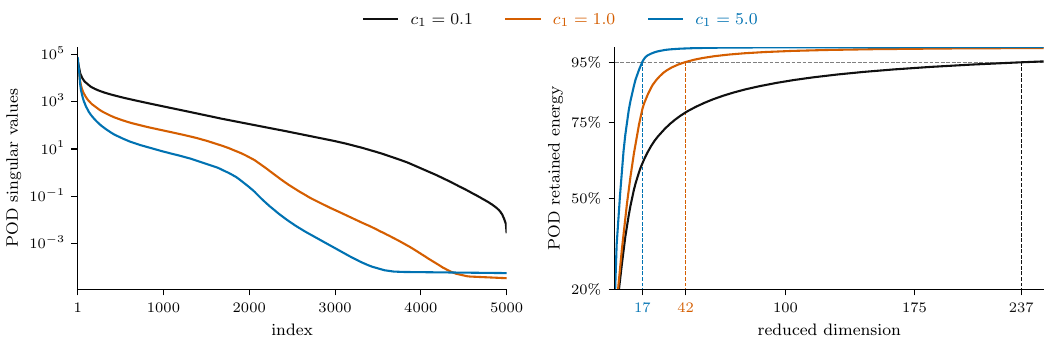}
    \caption{\label{fig:HW_pod_svals_and_ret_energy_c1_all}
        POD singular values (left) and corresponding retained energy (right) for $c_1 \in \{0.1, 1.0, 5.0\}$.
    }
\end{figure*}

We start by computing the three POD bases using the given $5,000$ DNS training snapshots.
Recall that we neither scale nor center the data before computing the bases.
Figure~\ref{fig:HW_pod_svals_and_ret_energy_c1_all} plots the POD singular values on the left and the corresponding retained energy on the right.
The decay rate of the singular values increases with $c_1$: for $c_1 = 0.1$, the singular values decay slowly, indicating that the dynamics are dominated by small-scale fluctuations; for $c_1 = 5.0$, larger-scale, slow-moving structures become more important and the singular values decay faster.
The number of POD modes necessary to retain $95 \%$ of the energy, for example, decreases from $r=237$ for $c_1 = 0.1$ to $r = 42$ for $c_1 = 1.0$ and $r = 17$ for $c_1 = 5.0$.

\begin{figure*}[htb]
    \centering
    \includegraphics{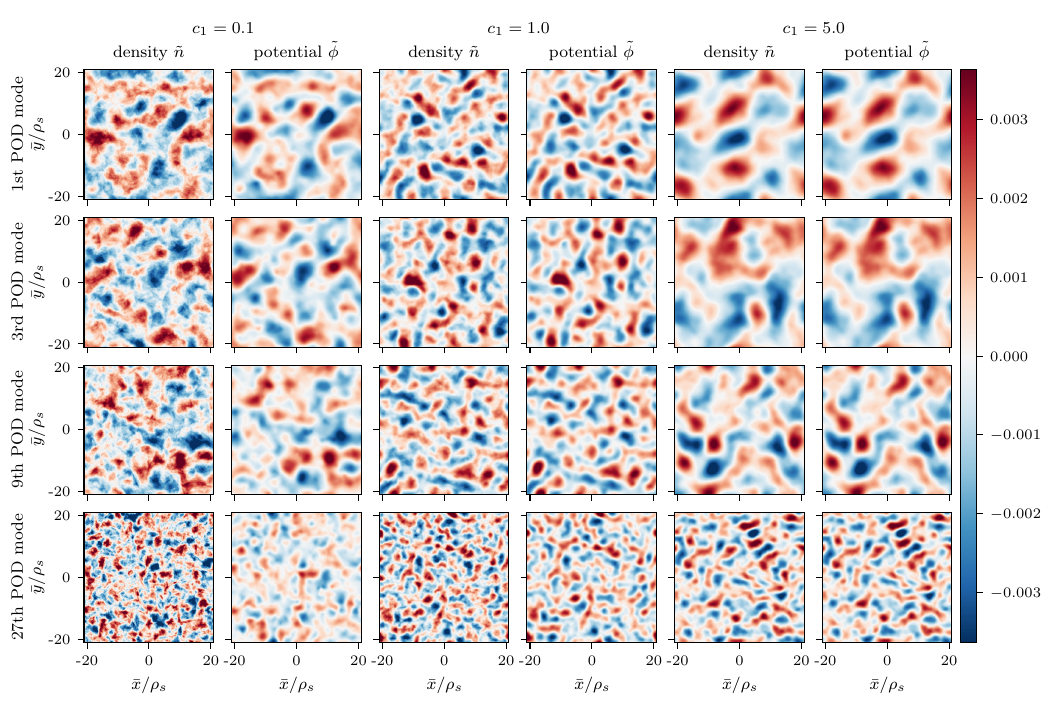}
    \caption{\label{fig:HW_POD_modes_all_c1}
        First, third, ninth, and 27th POD mode of the density and potential fluctuations for $c_1 \in \{ 0.1, 1.0, 5.0 \}$.
        The first modes retain $6.8 \%$, $7.5 \%$ and $17.6 \%$ of the total energy, respectively, while the 27th modes retain $0.7 \%$, $0.6 \%$ and $0.1 \%$ energy.
    }
\end{figure*}
Figure~\ref{fig:HW_POD_modes_all_c1} plots the first, third, ninth, and 27th POD mode of the density and potential fluctuations for the different $c_1$ values.
Note that POD sorts the modes in descending order based on their energy content, which means that the most energetic modes represent larger structures whereas the less energetic modes capture smaller-scale fluctuations.
For $c_1 \in \{ 0.1, 1.0, 5.0 \}$, the first modes retain $6.8 \%$, $7.5 \%$ and $17.6 \%$ of the energy while the 27th modes retain $0.7 \%$, $0.6 \%$ and $0.1 \%$ of the energy, respectively.
Notice that the similarity between the POD modes for density and potential fluctuations increases with $c_1$, which is consistent with the behavior of the corresponding fields (cf.\ Fig.~\ref{fig:reference_invariants_snapshots}, right).

The choice of the reduced dimension $r$ is essential to construct accurate and predictive OpInf ROMs.
Generally, $r$ should be chosen such that the corresponding POD modes retain as much energy as possible and, at the same time, it is not too large, to avoid learning an exceedingly large number of ROM coefficients via~\eqref{eq:rom_min_problem_state} and~\eqref{eq:rom_min_problem_output} which would be numerically challenging.
For $c_1 \in \{ 0.1, 1.0 \}$, where the dynamics are complex and dominated by small scale structures, we consider OpInf ROMs with reduced dimension $r=78$.
This value retains $86.22\%$ and $98.05\%$ of the total energy, respectively.
For $c_1 = 5.0$, we consider a lower reduced dimension, $r=44$, which retains $99.63 \%$ of the energy.
It is important to note that regardless of the considered reduced dimension, the least energetic POD modes $r + 1$ to $n_t$ will be neglected.
This, in turn, means that some of the finer-scale details will not be captured by the truncated basis and thus by the ensuing ROM.
In problems with turbulent dynamics such as the HW equations, these unresolved structures can be relevant for capturing small-scale turbulent effects.
In addition, there is usually a non-negligible interaction between the resolved and unresolved modes.
This means that in general, the ROM is able to model only a subset of the original dynamics, motivating the development of closure models~\cite{ahmedClosuresReducedOrder2021} as well as nonlinear approaches~\cite{brunaNeuralGalerkinSchemes2024,peherstorferModelReductionTransportDominated2020,gopakumarFourierNeuralOperator2023}.
We leave such investigations for our future research.

\subsection{Predictions obtained using Operator Inference}\label{subsec:opinf_results}
We construct OpInf ROMs following the steps presented in Sec.~\ref{sec:contributions} and use them for predictions for $400$ additional time units, that is, for a time horizon $400\%$ longer than the training horizon.
The learning of the reduced state operators is done via~\eqref{eq:rom_min_problem_state}.
However, since the scale of the projected snapshot data $\hat{\mathbf{Q}} \in \mathbb{R}^{r \times 5,000}$ computed using~\eqref{eq:opinf_projection} (which is exacerbated in $\hat{\mathbf{Q}} \otimes \hat{\mathbf{Q}}$) is different from the scales of the outputs in $\mathbf{Y} \in \mathbb{R}^{2 \times 5,000}$ for all three values of $c_1$, we first center (about the temporal mean) and scale (with respect to the maximum absolute value of the centered data) the projected snapshot data prior to inferring the reduced output operators.
This ensures that the transformed projected data do not exceed $[-1, 1]$, which is on similar scale as the outputs in $\mathbf{Y}$.
Centering introduces a constant and linear operator in the reduced output~\eqref{eq:opinf_quad_model_output} in addition to the original quadratic reduced operator.
The learning of the two additional operators is regularized via a hyperparameter $\beta_B > 0$ analogously to the reduced linear operator in Eq.~\eqref{eq:opinf_quad_model_state}.

To compute the ensemble of solutions, we use a grid of size $10 \times 10$ for the regularization parameters in the reduced state minimization problem~\eqref{eq:rom_min_problem_state} and a grid of size $20 \times 20$ for the regularization parameters to infer the (transformed) reduced output operators for each $c_1$.
We employ grids with a relatively large cardinality because we want to obtain a comprehensive assessment of the OpInf predictions.
Empirically, we observed that grid sizes between $5 \times 5$ and $20 \times 20$ suffice.
We use relative error thresholds of $5\%$ for the means and $30\%$ for the standard deviations of the two quantities of interest, $\Gamma_n$ and $\Gamma_c$.
In general, relative error thresholds between $5$ and $10\%$ for the means and between $15$ and $40 \%$ for the standard deviations retain models with the desired properties, namely long-time stability and statistical accuracy over the prediction horizon.
For a more in-depth analysis, we also extract the solution that minimizes the average relative error of both mean and standard deviation of $\Gamma_n$ and $\Gamma_c$ over the prediction time horizon.

Tables~\ref{table:HW_predictions_c1_0_1_train}~--~\ref{table:HW_predictions_c1_5_0_pred} show the means and standard deviations of the two quantities of interest over both the training and prediction horizons, for all $c_1$ values.
The first column in each table shows the reference results obtained from the DNS solution.
In the second column, we present the means and standard deviations corresponding to the average ensemble solutions, and in the third column, we display the results for the ensemble solution that minimizes the average relative error of both mean and standard deviation of $\Gamma_n$ and $\Gamma_c$ over the prediction horizon.
These results indicate that the OpInf ROMs achieve the primary objective and accurately match (up to the prescribed relative error tolerances) the reference means and standard deviations over both the training and prediction horizons.
More details including plots of the time traces of the two quantities of interest can be found in Appendix~\ref{appendix:time_traces_predictions}.
\begin{table}[h!]
    \caption{\label{table:HW_predictions_c1_0_1_train} Statistics of the OpInf output approximations for $c_1 = 0.1$ over the training horizon.}
    \begin{tabular}{c c c c c c}
                                     &      & ref. & ens.\ average & ens.\ min.\ err. & min.\ train err. \\
         \hline
         \multirow{2}{*}{$\Gamma_n$} & mean & 1.524 & 1.504          & 1.478          & 1.522 \\
                                     & std.  & 0.188 & 0.154          & 0.141          & 0.166 \\
         \multirow{2}{*}{$\Gamma_c$} & mean & 1.169 & 1.154          & 1.134          & 1.170 \\
                                     & std.  & 0.092 & 0.082          & 0.076          & 0.088
    \end{tabular}
\end{table}
\begin{table}[h!]
    \caption{\label{table:HW_predictions_c1_0_1_pred} Statistics of the OpInf output approximations for $c_1 = 0.1$ over the prediction horizon.}
    \begin{tabular}{c c c c c c}
                                     &      & ref. & ens.\ average & ens.\ min. err. & min.\ train error \\
         \hline
         \multirow{2}{*}{$\Gamma_n$} & mean & 1.513 & 1.538          & 1.541          & 1.559 \\
                                     & std.  & 0.211 & 0.168          & 0.183          & 0.054 \\
         \multirow{2}{*}{$\Gamma_c$} & mean & 1.168 & 1.181          & 1.182          & 1.227 \\
                                     & std.  & 0.122 & 0.088          & 0.126          & 0.030
    \end{tabular}
\end{table}
\begin{table}[h!]
    \caption{\label{table:HW_predictions_c1_1_0_train} Statistics of the OpInf output approximations for $c_1 = 1.0$ over the training horizon.}
    \begin{tabular}{c c c c c}
                                    &      & ref. & ens.\ average  & ens. min.  err. \\
        \hline
        \multirow{2}{*}{$\Gamma_n$} & mean & 0.624      & 0.616              & 0.620  \\
                                    & std.  & 0.042      & 0.038              & 0.038  \\
        \multirow{2}{*}{$\Gamma_c$} & mean & 0.616      & 0.608              & 0.612  \\
                                    & std.  & 0.041      & 0.037              & 0.038
    \end{tabular}
\end{table}
\begin{table}[h!]
    \caption{\label{table:HW_predictions_c1_1_0_pred} Statistics of the OpInf output approximations for $c_1 = 1.0$ over the prediction horizon.}
    \begin{tabular}{c c c c c}
                                    &      & ref. & ens. average  & ens.\ min.\ err. \\
        \hline
        \multirow{2}{*}{$\Gamma_n$} & mean & 0.607      & 0.621              & 0.619  \\
                                    & std.  & 0.047      & 0.047              & 0.039  \\
        \multirow{2}{*}{$\Gamma_c$} & mean & 0.605      & 0.609              & 0.607  \\
                                    & std.  & 0.041      & 0.034              & 0.022
    \end{tabular}
\end{table}
\begin{table}[h!]
    \caption{\label{table:HW_predictions_c1_5_0_train} Statistics of the OpInf output approximations for $c_1 = 5.0$ over the training horizon.}
    \begin{tabular}{c c c c c c}
                                     &      & ref. & ens.\ average  & ens.\ min.\ err. \\
         \hline
         \multirow{2}{*}{$\Gamma_n$} & mean & 0.130     & 0.130             & 0.129 \\
                                     & std.  & 0.009     & 0.009             & 0.008 \\
         \multirow{2}{*}{$\Gamma_c$} & mean & 0.124     & 0.124             & 0.122 \\
                                     & std.  & 0.010     & 0.010             & 0.009
    \end{tabular}
\end{table}
\begin{table}[h!]
    \caption{\label{table:HW_predictions_c1_5_0_pred} Statistics of the OpInf output approximations for $c_1 = 5.0$ over the prediction horizon.}
    \begin{tabular}{c c c c c c}
                                     &      & ref. & ens.\ average  & ens.\ min.\ err. \\
         \hline
         \multirow{2}{*}{$\Gamma_n$} & mean & 0.121     & 0.130             & 0.131 \\
                                     & std.  & 0.012     & 0.007             & 0.013 \\
         \multirow{2}{*}{$\Gamma_c$} & mean & 0.121     & 0.123             & 0.126 \\
                                     & std.  & 0.012     & 0.007             & 0.013
    \end{tabular}
\end{table}

\begin{figure*}
    \centering
    \includegraphics{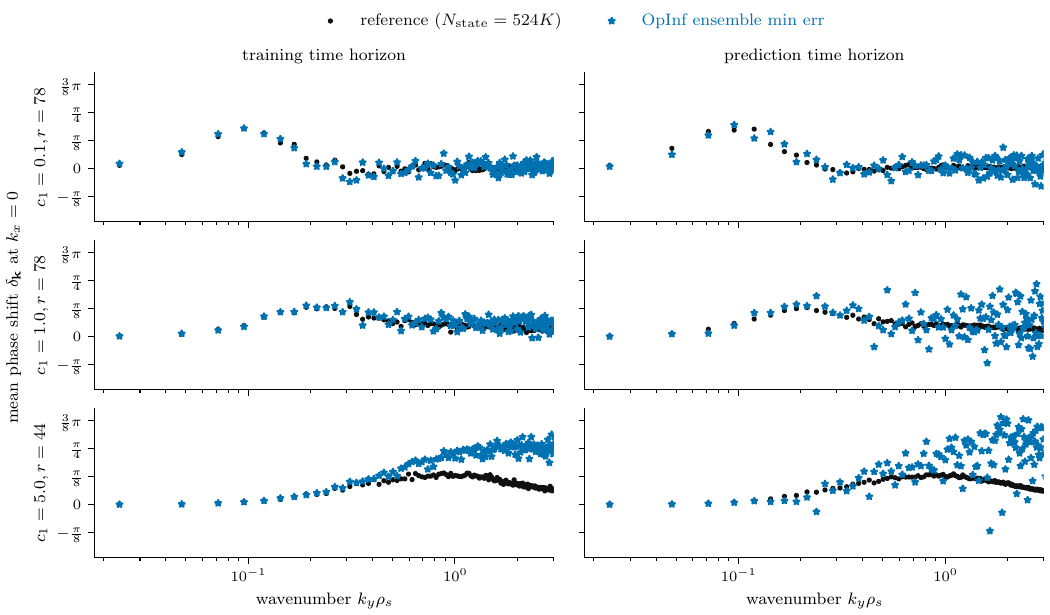}
    \caption{\label{fig:HW_phase_shift}
        The temporal mean of the phase shift $ \delta_\mathbf{k} $ at $ k_x = 0 $ over the training and prediction horizons.
    }
\end{figure*}

We next emphasize the importance of using regularization for obtaining stable and statistically accurate predictions with OpInf.
Without regularization, the OpInf ROMs diverge after roughly $10$ time iterations, i.e., $0.1$ time units, for all $c_1$ values.
Additionally, employing commonly used regularization strategies (for example, strategies that minimize the training error only) typically lead to inaccurate approximations.
We illustrate this point for $c_1 = 0.1$; a similar behavior was also observed for $c_1 = 1.0$ and $c_1 = 5.0$.
The last columns in Tables~\ref{table:HW_predictions_c1_0_1_train} and~\ref{table:HW_predictions_c1_0_1_pred} show the results obtained with an OpInf ROM constructed using a na\"ive regularization strategy, in which the optimal regularization parameters minimize the mean relative $L_2$ error for $\Gamma_n$ and $\Gamma_c$ over the training horizon, excluding diverging solutions.
Using the same search grid as in the other experiments, we find that the obtained ROM fails to capture the dynamics of the quantities of interest beyond training and only accurately captures their mean.
A similar behavior has been previously observed in~\cite{ghazijahaniPredictionTurbulentFlow2024}.

For a more in-depth assessment of the approximate state solutions obtained with our ROMs, Fig.~\ref{fig:HW_phase_shift} plots the temporal mean of the cross phases $ \delta_\mathbf{k}$~\eqref{eq:phase_shift} at $k_x = 0 $ for the ensemble solutions with minimal average relative errors.
It has been observed that the phase shift peaks at higher wavenumbers as $c_1$ increases~\cite{kimNumericalInvestigationFrequency2013}, which is consistent with what we observe in Fig.~\ref{fig:HW_phase_shift} as well.
The OpInf approximations accurately match low wavenumbers with $ k_y \rho_s \lesssim 0.4 $ over both training and prediction time horizon.
Smaller scales beyond $ k_y \gtrsim 0.4 $ are more scattered, with increasing amplitudes for higher adiabaticities, particularly for the predictions beyond the training time horizon.
This is consistent with the fact that due to truncation, the underlying POD bases cannot resolve some of the small-scale features corresponding to high wavenumber transport.
For $c_1 \in \{0.1, 1.0\}$, the predictions match the peaks of the reference phase shifts, particularly over the prediction time horizon.
This emphasizes that the corresponding ROMs correctly predict the bulk of underlying transport beyond training and accurately model the adiabatic response.
Conversely, for $ c_1 = 5.0 $, the OpInf approximations match the phase shift for low wavenumbers but fail to match the peak for both training and prediction time horizons.
The fact that the phase shift of the ROM prediction overshoots the peak over training suggests that the considered reduced dimension $r = 44$ might not be large enough.
More detailed investigations indeed show that we would need a large reduced dimension exceeding $500$ to accurately resolve the peak phase shift for $c_1 = 5.0 $, which would be prohibitive.
We also note that for $c_1 = 5.0 $, the system is strongly adiabatic, and the fields are very similar (cf.\ Fig.~\ref{fig:reference_invariants_snapshots}, right).
Thus, the phase shift alone might not be a reliable indicator of accurate modeled adiabaticity, as it does not consider the relative amplitudes of the fields~\cite{camargoResistiveDriftWave1995}.

\begin{figure}
    \centering
    \includegraphics{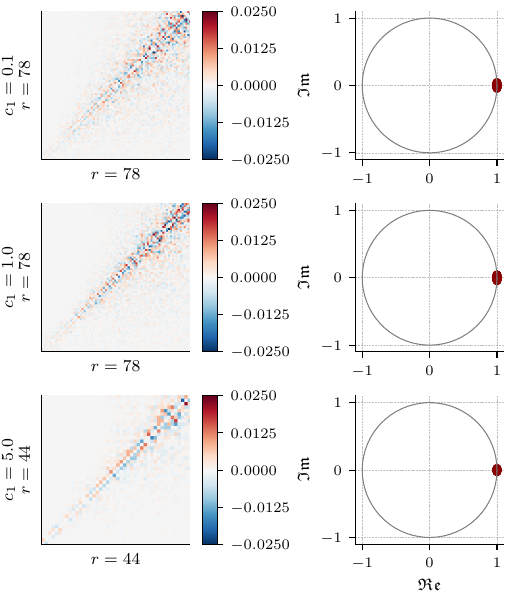}
    \caption{\label{fig:HW_spectrum_red_lin_op}
        Off-diagonal entries of the reduced linear state operators (left) and the corresponding spectra (right).
    }
\end{figure}

Finally, we take a closer look at the inferred reduced linear state operators, $\hat{\mathbf{A}} \in \mathbb{R}^{r \times r}$, corresponding to the ensemble solution that minimizes the average prediction error.
Figure~\ref{fig:HW_spectrum_red_lin_op} plots the reduced operators off-diagonal entries on the left and their spectra on the right.
All three reduced operators have diagonal entries close to $1.0$ and off-diagonal elements exhibiting a quasi-band structure with magnitudes smaller than $1.0$ that decay as the distance from the diagonal increases.
The eigenvalues of the reduced linear operators lie on or very close to the unit circle in the stability plane.
We note that all ensemble solutions exhibited this behavior.
Discarded solutions included reduced linear state operators with unstable modes (these solutions typically diverged) as well as strongly damped spectra obtained with large regularization parameters.
This positions the spectra at the threshold of linear stability, which allows the linear reduced operators to approximate the oscillatory behavior of the reference solutions.
Compared to the spectrum of the linear operator $\mathbf{A}$ of the HW equations, the originally linearly unstable modes are dampened due to POD basis truncation and regularization in the corresponding OpInf learning problem.
This motivates further exploration into incorporating closure models into our approach and nonlinear methods.

The runtimes of the OpInf ROMs with reduced dimension $r=78$, averaged over $100$ runs, are $7.024 \pm 0.07$ seconds for computing the reduced state via~\eqref{eq:opinf_quad_model_state} and $0.027 \pm 0.0038$ seconds for estimating the output using~\eqref{eq:opinf_quad_model_output}.
The corresponding runtimes for the ROM with $r=44$ are $3.55 \pm 0.047$ and $0.0048 \pm 0.002$ seconds, respectively.
This translates into four and respectively five orders of magnitude improvement in single-core performance compared the high-fidelity simulations.
Solving the ROM and predicting future dynamics is therefore computationally cheap and can be done close to real-time on laptop computers or similar devices.

\subsection{The effect of the initial condition used to generate the training data on the reduced model predictions}\label{subsec:opinf_sensitivity}

We close this section with an important remark regarding the effect of the initial conditions used to generate the high-fidelity training data on OpInf ROMs.
While different random initial conditions in the HW system will generate statistically similar outputs over the training and prediction horizons, their local features and trends can diverge significantly due to the system's chaotic nature.
The discussion in Appendix~\ref{appendix:time_traces_predictions} highlights this point: the OpInf ROMs accurately capture the statistical properties of the two outputs, but because the reference solutions exhibit differing local features between the training and prediction horizons, the ROM predictions are less effective in capturing this structural change.
Using a different training dataset where the trends over the training and prediction horizons are more similar would decrease the mismatch between the trends of the reference and ROM solutions.
We emphasize this point by performing an additional experiment.

The training data in the new experiment are generated for $c_1 = 1.0$ over the time horizon $[0; 1,000]$ using $\delta_t = 0.025$.
To ensure that the turbulent dynamics are well-developed, we consider $t_{\mathrm{init}} = 500$.
We then train an OpInf ROM using data for $100$ time units with reduced dimension $r=44$, retaining $95\%$ of the total energy.
As in the previous experiments, we use the ROM for predictions for $400$ additional time units.
Further, we use a relative error threshold of $5\%$ for the means and $15\%$ for the standard deviations.

Tables~\ref{table:HW_predictions_c1_1_0_train_new} and~\ref{table:HW_predictions_c1_1_0_pred_new} compare the statistics of the OpInf ensemble average solution and the ensemble solution that minimizes the relative errors over the prediction horizon with the reference solution.
We see that the statistics of the ROM approximations accurately match the reference solutions.
The similarity between the trends of the reference solution over the training and prediction horizons was crucial for obtaining these results.
This is further emphasized in Fig.~\ref{fig:HW_predictions_c1_1_0_nt_4000} in Appendix~\ref{appendix:opinf_sensitivity}, which show that the OpInf ROM approximate solutions exhibit a closer alignment with the reference solutions over the prediction horizon.

\begin{table}[htb]
    \caption{\label{table:HW_predictions_c1_1_0_train_new} Statistics of the OpInf output approximations for $c_1 = 1.0$ over the training horizon when the training dataset was generated using a different initial condition for the DNS.}
    \begin{tabular}{c c c c c}
                                    &      & ref. & ens.\ average  & ens.\ min.\ err. \\
        \hline
        \multirow{2}{*}{$\Gamma_n$} & mean & 0.627      & 0.628              & 0.627  \\
                                    & std.  & 0.045      & 0.044              & 0.044  \\
        \multirow{2}{*}{$\Gamma_c$} & mean & 0.618      & 0.619              & 0.618  \\
                                    & std.  & 0.041      & 0.039              & 0.039
    \end{tabular}
\end{table}
\begin{table}[htb]
    \caption{\label{table:HW_predictions_c1_1_0_pred_new} Statistics of the OpInf output approximations for $c_1 = 1.0$ over the prediction horizon when the training dataset was generated using a different initial condition for the DNS.}
    \begin{tabular}{c c c c c}
                                    &      & ref. & ens.\ average  & ens.\ min.\ err. \\
        \hline
        \multirow{2}{*}{$\Gamma_n$} & mean & 0.638      & 0.651              & 0.621  \\
                                    & std.  & 0.041      & 0.044              & 0.046  \\
        \multirow{2}{*}{$\Gamma_c$} & mean & 0.629      & 0.628              & 0.604  \\
                                    & std.  & 0.035      & 0.035              & 0.035
    \end{tabular}
\end{table}

Two strategies can be employed to mitigate the influence of a specific initial condition on the long-term behavior of the ROM and enhance its robustness: extending the training horizon or utilizing a training dataset that incorporates trajectories from multiple initial conditions.
Nonetheless, both approaches come with an increased computational cost for generating the training data.

\section{Summary and conclusions}\label{sec:conclusions}
In this paper, we showed that Scientific Machine Learning has the potential to construct predictive reduced models from data for complex plasma turbulence models such as the (classical) Hasegawa-Wakatani equations.
To this end, we investigated using Operator Inference.
This approach is flexible and easy to use, and only requires knowledge about the structure of the underlying governing equations and data to train a structure-preserving reduced model.
Our results showed that Operator Inference can be employed to construct reduced models that provide stable and statistically accurate approximations over long time horizons that can also capture some of the important features of the turbulent dynamics while reducing the computational effort by orders of magnitude.
This in turn enables performing these simulations on laptop computers or similar devices instead of using compute clusters or supercomputers.
In the broader context of fusion research, constructing such reduced models can have a significant impact and amount to a milestone toward real-time plasma turbulence simulations as well as the design of optimized fusion devices.
While these results illustrate the potential of the approach, they also highlight a number of open questions.
Important future directions include the investigation of closure models to account for both the truncation of the reduced basis and the interaction between the resolved and unresolved modes.
An alternative direction involves investigating nonlinear approaches for model reduction~\cite{gopakumarFourierNeuralOperator2023, brunaNeuralGalerkinSchemes2024,peherstorferModelReductionTransportDominated2020}, which were shown to be effective in problems with complex dynamics.

\begin{acknowledgments}
    The authors gratefully acknowledge Robin Greif and Andreas Stegmeir for fruitful discussions.
    C.~Gahr was supported by the Helmholtz Association under the joint research school ``Munich School for Data Science -- MUDS''.
    Computations were performed on the HPC system Cobra and Raven at the Max Planck Computing and Data Facility.
\end{acknowledgments}

\section*{Data availability statement}
The high-fidelity simulation data were generated using the \texttt{hw2d} code~\cite{greifHW2DReferenceImplementation2023}.
The code and data to reproduce the results of this paper are available at \url{https://gitlab.mpcdf.mpg.de/cgahr/learning-roms-for-hw}.

\appendix
\section{Time traces of the quantities of interest}\label{appendix:time_traces_predictions}

\begin{figure*}[p]
    \centering
    \includegraphics{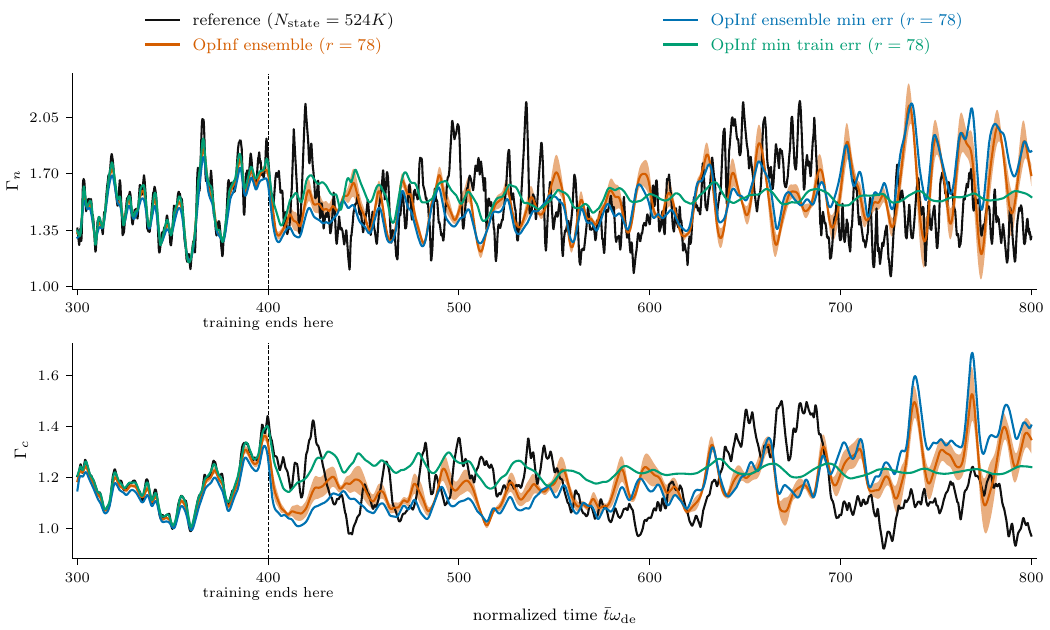}
    \caption{\label{fig:HW_predictions_c1_0_1}
        OpInf output approximations for $c_1 = 0.1$.
    }
\end{figure*}

\begin{figure*}[p]
    \centering
    \includegraphics{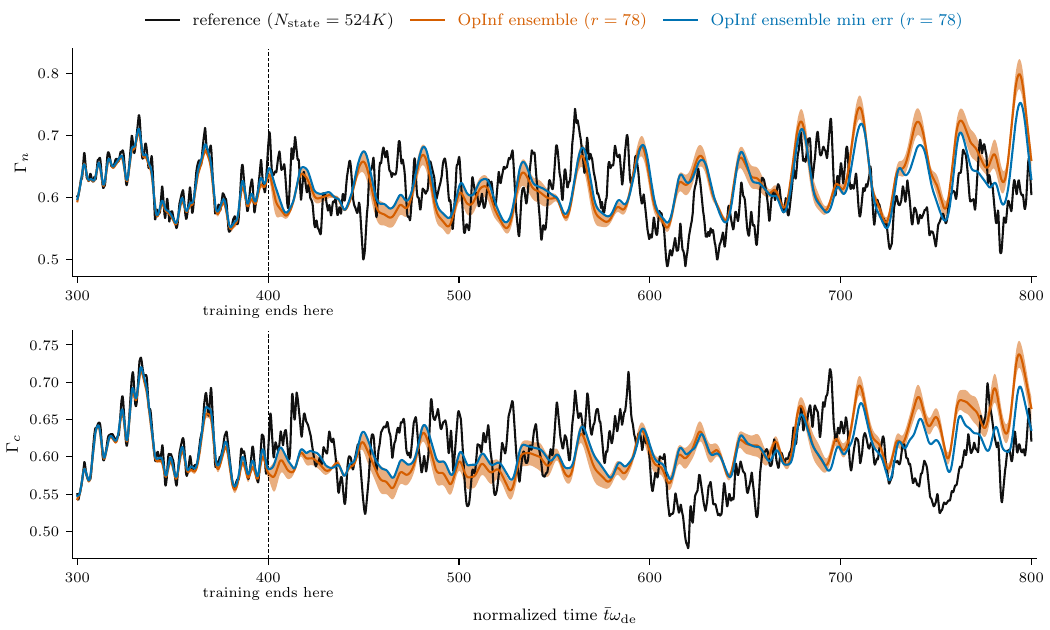}
    \caption{\label{fig:HW_predictions_c1_1_0}
        OpInf output approximations for $c_1 = 1.0$.
    }
\end{figure*}

\begin{figure*}[p]
    \centering
    \includegraphics{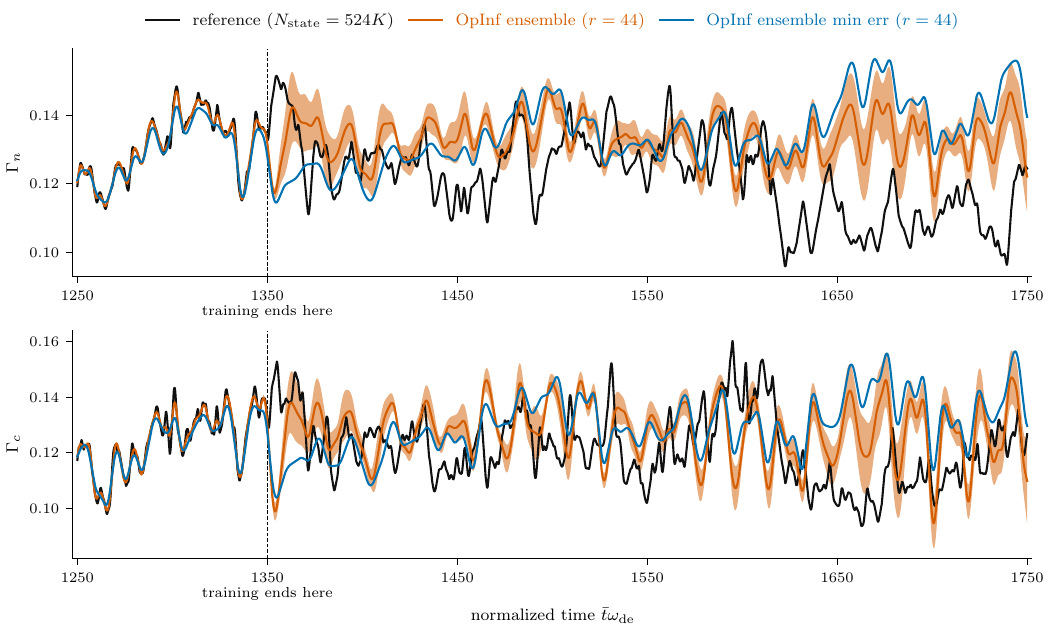}
    \caption{\label{fig:HW_predictions_c1_5_0}
        OpInf output approximations for $c_1 = 5.0$.
    }
\end{figure*}

Figs.~\ref{fig:HW_predictions_c1_0_1}--\ref{fig:HW_predictions_c1_5_0} plot the OpInf approximations for $\Gamma_n$ and $\Gamma_c$.
In orange, we show the ensemble mean and one standard deviation of the OpInf output approximations.
In blue, we plot the ensemble solution that minimizes the relative errors of the statistics.
The main purpose of these figures is to provide a quantitative assessment of the ROM approximate solutions.
As stated in the main text, due to the nonlinear and chaotic nature of the HW system, our ROMs are not expected to provide point-wise accurate predictions but rather predictions that are statistically accurate.
Qualitatively, the OpInf approximations capture some of the overall trends while finer details are not well captured.
This is expected due to the truncation of the POD basis vectors.
Additionally, it is important to note that the local features differ considerably between training and prediction horizons due to the chaotic nature of the problem.
Consequently, OpInf ROMs inherently cannot capture features that were not present in the training dataset and different initial conditions used to generate the training data can impact the accuracy of the reduced models predictions.


\section{Time traces of the quantities of interest obtained using a different realization of the initial condition for \texorpdfstring{$c_1 = 1.0$}{c1 = 1.0}}\label{appendix:opinf_sensitivity}

\begin{figure*}[p]
    \centering
    \includegraphics{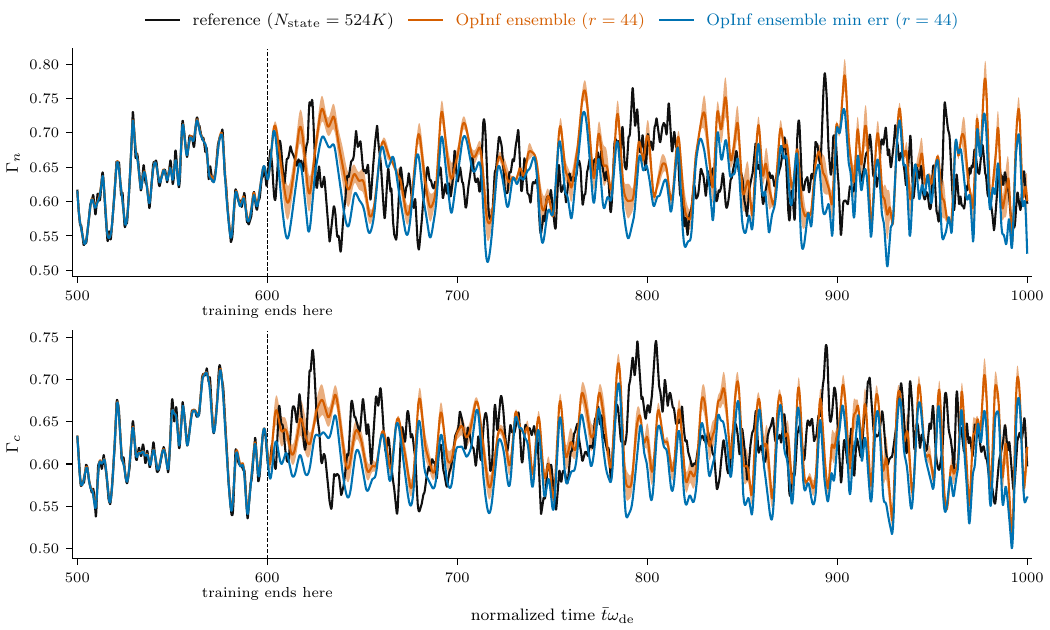}
    \caption{\label{fig:HW_predictions_c1_1_0_nt_4000}
        OpInf output approximations for $c_1 = 1.0$ when the training dataset was generated using a different initial condition for the DNS.\@
    }
\end{figure*}

Figure~\ref{fig:HW_predictions_c1_1_0_nt_4000} show that the OpInf ROM approximations accurately match the reference statistics and also capture the trends of the reference solutions over the prediction horizon.
The similarity between the trends of the reference solution over the training and prediction horizons was crucial for obtaining these results.

\clearpage
\bibliography{SciML_ROM_HW}

\begin{thebibliography}{59}%
\makeatletter
\providecommand \@ifxundefined [1]{%
 \@ifx{#1\undefined}
}%
\providecommand \@ifnum [1]{%
 \ifnum #1\expandafter \@firstoftwo
 \else \expandafter \@secondoftwo
 \fi
}%
\providecommand \@ifx [1]{%
 \ifx #1\expandafter \@firstoftwo
 \else \expandafter \@secondoftwo
 \fi
}%
\providecommand \natexlab [1]{#1}%
\providecommand \enquote  [1]{``#1''}%
\providecommand \bibnamefont  [1]{#1}%
\providecommand \bibfnamefont [1]{#1}%
\providecommand \citenamefont [1]{#1}%
\providecommand \href@noop [0]{\@secondoftwo}%
\providecommand \href [0]{\begingroup \@sanitize@url \@href}%
\providecommand \@href[1]{\@@startlink{#1}\@@href}%
\providecommand \@@href[1]{\endgroup#1\@@endlink}%
\providecommand \@sanitize@url [0]{\catcode `\\12\catcode `\$12\catcode
  `\&12\catcode `\#12\catcode `\^12\catcode `\_12\catcode `\%12\relax}%
\providecommand \@@startlink[1]{}%
\providecommand \@@endlink[0]{}%
\providecommand \url  [0]{\begingroup\@sanitize@url \@url }%
\providecommand \@url [1]{\endgroup\@href {#1}{\urlprefix }}%
\providecommand \urlprefix  [0]{URL }%
\providecommand \Eprint [0]{\href }%
\providecommand \doibase [0]{http://dx.doi.org/}%
\providecommand \selectlanguage [0]{\@gobble}%
\providecommand \bibinfo  [0]{\@secondoftwo}%
\providecommand \bibfield  [0]{\@secondoftwo}%
\providecommand \translation [1]{[#1]}%
\providecommand \BibitemOpen [0]{}%
\providecommand \bibitemStop [0]{}%
\providecommand \bibitemNoStop [0]{.\EOS\space}%
\providecommand \EOS [0]{\spacefactor3000\relax}%
\providecommand \BibitemShut  [1]{\csname bibitem#1\endcsname}%
\let\auto@bib@innerbib\@empty
\bibitem [{\citenamefont {Atchley}\ \emph {et~al.}(2023)\citenamefont
  {Atchley}, \citenamefont {Zimmer}, \citenamefont {Lange}, \citenamefont
  {Bernholdt}, \citenamefont {Melesse~Vergara}, \citenamefont {Beck},
  \citenamefont {Brim}, \citenamefont {Budiardja}, \citenamefont
  {Chandrasekaran}, \citenamefont {Eisenbach}, \citenamefont {Evans},
  \citenamefont {Ezell}, \citenamefont {Frontiere}, \citenamefont {Georgiadou},
  \citenamefont {Glenski}, \citenamefont {Grete}, \citenamefont {Hamilton},
  \citenamefont {Holmen}, \citenamefont {Huebl}, \citenamefont {Jacobson},
  \citenamefont {Joubert}, \citenamefont {Mcmahon}, \citenamefont {Merzari},
  \citenamefont {Moore}, \citenamefont {Myers}, \citenamefont {Nichols},
  \citenamefont {Oral}, \citenamefont {Papatheodore}, \citenamefont {Perez},
  \citenamefont {Rogers}, \citenamefont {Schneider}, \citenamefont {Vay},\ and\
  \citenamefont {Yeung}}]{atchleyFrontierExploringExascale2023}%
  \BibitemOpen
  \bibfield  {author} {\bibinfo {author} {\bibfnamefont {S.}~\bibnamefont
  {Atchley}}, \bibinfo {author} {\bibfnamefont {C.}~\bibnamefont {Zimmer}},
  \bibinfo {author} {\bibfnamefont {J.}~\bibnamefont {Lange}}, \bibinfo
  {author} {\bibfnamefont {D.}~\bibnamefont {Bernholdt}}, \bibinfo {author}
  {\bibfnamefont {V.}~\bibnamefont {Melesse~Vergara}}, \bibinfo {author}
  {\bibfnamefont {T.}~\bibnamefont {Beck}}, \bibinfo {author} {\bibfnamefont
  {M.}~\bibnamefont {Brim}}, \bibinfo {author} {\bibfnamefont {R.}~\bibnamefont
  {Budiardja}}, \bibinfo {author} {\bibfnamefont {S.}~\bibnamefont
  {Chandrasekaran}}, \bibinfo {author} {\bibfnamefont {M.}~\bibnamefont
  {Eisenbach}}, \bibinfo {author} {\bibfnamefont {T.}~\bibnamefont {Evans}},
  \bibinfo {author} {\bibfnamefont {M.}~\bibnamefont {Ezell}}, \bibinfo
  {author} {\bibfnamefont {N.}~\bibnamefont {Frontiere}}, \bibinfo {author}
  {\bibfnamefont {A.}~\bibnamefont {Georgiadou}}, \bibinfo {author}
  {\bibfnamefont {J.}~\bibnamefont {Glenski}}, \bibinfo {author} {\bibfnamefont
  {P.}~\bibnamefont {Grete}}, \bibinfo {author} {\bibfnamefont
  {S.}~\bibnamefont {Hamilton}}, \bibinfo {author} {\bibfnamefont
  {J.}~\bibnamefont {Holmen}}, \bibinfo {author} {\bibfnamefont
  {A.}~\bibnamefont {Huebl}}, \bibinfo {author} {\bibfnamefont
  {D.}~\bibnamefont {Jacobson}}, \bibinfo {author} {\bibfnamefont
  {W.}~\bibnamefont {Joubert}}, \bibinfo {author} {\bibfnamefont
  {K.}~\bibnamefont {Mcmahon}}, \bibinfo {author} {\bibfnamefont
  {E.}~\bibnamefont {Merzari}}, \bibinfo {author} {\bibfnamefont
  {S.}~\bibnamefont {Moore}}, \bibinfo {author} {\bibfnamefont
  {A.}~\bibnamefont {Myers}}, \bibinfo {author} {\bibfnamefont
  {S.}~\bibnamefont {Nichols}}, \bibinfo {author} {\bibfnamefont
  {S.}~\bibnamefont {Oral}}, \bibinfo {author} {\bibfnamefont {T.}~\bibnamefont
  {Papatheodore}}, \bibinfo {author} {\bibfnamefont {D.}~\bibnamefont {Perez}},
  \bibinfo {author} {\bibfnamefont {D.~M.}\ \bibnamefont {Rogers}}, \bibinfo
  {author} {\bibfnamefont {E.}~\bibnamefont {Schneider}}, \bibinfo {author}
  {\bibfnamefont {J.-L.}\ \bibnamefont {Vay}}, \ and\ \bibinfo {author}
  {\bibfnamefont {P.~K.}\ \bibnamefont {Yeung}},\ }in\ \href {\doibase
  10.1145/3581784.3607089} {\emph {\bibinfo {booktitle} {Proceedings of the
  {{International Conference}} for {{High Performance Computing}},
  {{Networking}}, {{Storage}} and {{Analysis}}}}},\ \bibinfo {series and
  number} {{{SC}} '23}\ (\bibinfo  {publisher} {Association for Computing
  Machinery},\ \bibinfo {address} {New York, NY, USA},\ \bibinfo {year}
  {2023})\ pp.\ \bibinfo {pages} {1--16}\BibitemShut {NoStop}%
\bibitem [{\citenamefont {Germaschewski}\ \emph {et~al.}(2021)\citenamefont
  {Germaschewski}, \citenamefont {Allen}, \citenamefont {Dannert},
  \citenamefont {Hrywniak}, \citenamefont {Donaghy}, \citenamefont {Merlo},
  \citenamefont {Ethier}, \citenamefont {D'Azevedo}, \citenamefont {Jenko},\
  and\ \citenamefont
  {Bhattacharjee}}]{germaschewskiExascaleWholedeviceModeling2021}%
  \BibitemOpen
  \bibfield  {author} {\bibinfo {author} {\bibfnamefont {K.}~\bibnamefont
  {Germaschewski}}, \bibinfo {author} {\bibfnamefont {B.}~\bibnamefont
  {Allen}}, \bibinfo {author} {\bibfnamefont {T.}~\bibnamefont {Dannert}},
  \bibinfo {author} {\bibfnamefont {M.}~\bibnamefont {Hrywniak}}, \bibinfo
  {author} {\bibfnamefont {J.}~\bibnamefont {Donaghy}}, \bibinfo {author}
  {\bibfnamefont {G.}~\bibnamefont {Merlo}}, \bibinfo {author} {\bibfnamefont
  {S.}~\bibnamefont {Ethier}}, \bibinfo {author} {\bibfnamefont
  {E.}~\bibnamefont {D'Azevedo}}, \bibinfo {author} {\bibfnamefont
  {F.}~\bibnamefont {Jenko}}, \ and\ \bibinfo {author} {\bibfnamefont
  {A.}~\bibnamefont {Bhattacharjee}},\ }\href {\doibase 10.1063/5.0046327}
  {\bibfield  {journal} {\bibinfo  {journal} {Physics of Plasmas}\ }\textbf
  {\bibinfo {volume} {28}},\ \bibinfo {pages} {062501} (\bibinfo {year}
  {2021})}\BibitemShut {NoStop}%
\bibitem [{\citenamefont {Belli}, \citenamefont {Candy},\ and\ \citenamefont
  {Sfiligoi}(2022)}]{belliSpectralTransitionMultiscale2022}%
  \BibitemOpen
  \bibfield  {author} {\bibinfo {author} {\bibfnamefont {E.~A.}\ \bibnamefont
  {Belli}}, \bibinfo {author} {\bibfnamefont {J.}~\bibnamefont {Candy}}, \ and\
  \bibinfo {author} {\bibfnamefont {I.}~\bibnamefont {Sfiligoi}},\ }\href@noop
  {} {\bibfield  {journal} {\bibinfo  {journal} {Plasma Physics and Controlled
  Fusion}\ }\textbf {\bibinfo {volume} {65}},\ \bibinfo {pages} {024001}
  (\bibinfo {year} {2022})}\BibitemShut {NoStop}%
\bibitem [{\citenamefont {Farca{\c s}}, \citenamefont {Merlo},\ and\
  \citenamefont {Jenko}(2022)}]{farcasGeneralFrameworkQuantifying2022}%
  \BibitemOpen
  \bibfield  {author} {\bibinfo {author} {\bibfnamefont {I.-G.}\ \bibnamefont
  {Farca{\c s}}}, \bibinfo {author} {\bibfnamefont {G.}~\bibnamefont {Merlo}},
  \ and\ \bibinfo {author} {\bibfnamefont {F.}~\bibnamefont {Jenko}},\
  }\href@noop {} {\bibfield  {journal} {\bibinfo  {journal} {Communications
  Engineering}\ }\textbf {\bibinfo {volume} {1}},\ \bibinfo {pages} {43}
  (\bibinfo {year} {2022})}\BibitemShut {NoStop}%
\bibitem [{\citenamefont {Benner}\ \emph {et~al.}(2017)\citenamefont {Benner},
  \citenamefont {Ohlberger}, \citenamefont {Cohen},\ and\ \citenamefont
  {Willcox}}]{bennerModelReductionApproximation2017}%
  \BibitemOpen
  \bibfield  {author} {\bibinfo {author} {\bibfnamefont {P.}~\bibnamefont
  {Benner}}, \bibinfo {author} {\bibfnamefont {M.}~\bibnamefont {Ohlberger}},
  \bibinfo {author} {\bibfnamefont {A.}~\bibnamefont {Cohen}}, \ and\ \bibinfo
  {author} {\bibfnamefont {K.}~\bibnamefont {Willcox}},\ }\href {\doibase
  10.1137/1.9781611974829} {\emph {\bibinfo {title} {Model {{Reduction}} and
  {{Approximation}}}}},\ Computational {{Science}} \& {{Engineering}}\
  (\bibinfo  {publisher} {{Society for Industrial and Applied Mathematics}},\
  \bibinfo {year} {2017})\BibitemShut {NoStop}%
\bibitem [{\citenamefont {Benner}, \citenamefont {Gugercin},\ and\
  \citenamefont {Willcox}(2015)}]{bennerSurveyProjectionBasedModel2015}%
  \BibitemOpen
  \bibfield  {author} {\bibinfo {author} {\bibfnamefont {P.}~\bibnamefont
  {Benner}}, \bibinfo {author} {\bibfnamefont {S.}~\bibnamefont {Gugercin}}, \
  and\ \bibinfo {author} {\bibfnamefont {K.}~\bibnamefont {Willcox}},\ }\href
  {\doibase 10.1137/130932715} {\bibfield  {journal} {\bibinfo  {journal} {SIAM
  Review}\ }\textbf {\bibinfo {volume} {57}},\ \bibinfo {pages} {483} (\bibinfo
  {year} {2015})}\BibitemShut {NoStop}%
\bibitem [{\citenamefont {Peherstorfer}\ and\ \citenamefont
  {Willcox}(2016)}]{peherstorferDatadrivenOperatorInference2016}%
  \BibitemOpen
  \bibfield  {author} {\bibinfo {author} {\bibfnamefont {B.}~\bibnamefont
  {Peherstorfer}}\ and\ \bibinfo {author} {\bibfnamefont {K.}~\bibnamefont
  {Willcox}},\ }\href {\doibase 10.1016/j.cma.2016.03.025} {\bibfield
  {journal} {\bibinfo  {journal} {Computer Methods in Applied Mechanics and
  Engineering}\ }\textbf {\bibinfo {volume} {306}},\ \bibinfo {pages} {196}
  (\bibinfo {year} {2016})}\BibitemShut {NoStop}%
\bibitem [{\citenamefont {Kramer}, \citenamefont {Peherstorfer},\ and\
  \citenamefont {Willcox}(2024)}]{kramerLearningNonlinearReduced2024}%
  \BibitemOpen
  \bibfield  {author} {\bibinfo {author} {\bibfnamefont {B.}~\bibnamefont
  {Kramer}}, \bibinfo {author} {\bibfnamefont {B.}~\bibnamefont
  {Peherstorfer}}, \ and\ \bibinfo {author} {\bibfnamefont {K.~E.}\
  \bibnamefont {Willcox}},\ }\href {\doibase
  10.1146/annurev-fluid-121021-025220} {\bibfield  {journal} {\bibinfo
  {journal} {Annual Review of Fluid Mechanics}\ }\textbf {\bibinfo {volume}
  {56}},\ \bibinfo {pages} {521} (\bibinfo {year} {2024})}\BibitemShut
  {NoStop}%
\bibitem [{\citenamefont {Kramer}\ and\ \citenamefont
  {Willcox}(2019)}]{kramerNonlinearModelOrder2019}%
  \BibitemOpen
  \bibfield  {author} {\bibinfo {author} {\bibfnamefont {B.}~\bibnamefont
  {Kramer}}\ and\ \bibinfo {author} {\bibfnamefont {K.~E.}\ \bibnamefont
  {Willcox}},\ }\href {\doibase 10.2514/1.J057791} {\bibfield  {journal}
  {\bibinfo  {journal} {AIAA Journal}\ }\textbf {\bibinfo {volume} {57}},\
  \bibinfo {pages} {2297} (\bibinfo {year} {2019})}\BibitemShut {NoStop}%
\bibitem [{\citenamefont {Qian}\ \emph {et~al.}(2020)\citenamefont {Qian},
  \citenamefont {Kramer}, \citenamefont {Peherstorfer},\ and\ \citenamefont
  {Willcox}}]{qianLiftLearnPhysicsinformed2020}%
  \BibitemOpen
  \bibfield  {author} {\bibinfo {author} {\bibfnamefont {E.}~\bibnamefont
  {Qian}}, \bibinfo {author} {\bibfnamefont {B.}~\bibnamefont {Kramer}},
  \bibinfo {author} {\bibfnamefont {B.}~\bibnamefont {Peherstorfer}}, \ and\
  \bibinfo {author} {\bibfnamefont {K.}~\bibnamefont {Willcox}},\ }\href
  {\doibase 10.1016/j.physd.2020.132401} {\bibfield  {journal} {\bibinfo
  {journal} {Physica D: Nonlinear Phenomena}\ }\textbf {\bibinfo {volume}
  {406}},\ \bibinfo {pages} {132401} (\bibinfo {year} {2020})}\BibitemShut
  {NoStop}%
\bibitem [{\citenamefont {Farcas}\ \emph
  {et~al.}(2024{\natexlab{a}})\citenamefont {Farcas}, \citenamefont {Gundevia},
  \citenamefont {Munipalli},\ and\ \citenamefont
  {Willcox}}]{farcasDistributedComputingPhysicsbased2024}%
  \BibitemOpen
  \bibfield  {author} {\bibinfo {author} {\bibfnamefont {I.-G.}\ \bibnamefont
  {Farcas}}, \bibinfo {author} {\bibfnamefont {R.~P.}\ \bibnamefont
  {Gundevia}}, \bibinfo {author} {\bibfnamefont {R.}~\bibnamefont {Munipalli}},
  \ and\ \bibinfo {author} {\bibfnamefont {K.~E.}\ \bibnamefont {Willcox}},\
  }\href@noop {} {\enquote {\bibinfo {title} {Distributed computing for
  physics-based data-driven reduced modeling at scale: {{Application}} to a
  rotating detonation rocket engine},}\ } (\bibinfo {year}
  {2024}{\natexlab{a}}),\ \Eprint {http://arxiv.org/abs/2407.09994}
  {arXiv:2407.09994 [cs, math]} \BibitemShut {NoStop}%
\bibitem [{\citenamefont {Farcas}\ \emph {et~al.}(2023)\citenamefont {Farcas},
  \citenamefont {Gundevia}, \citenamefont {Munipalli},\ and\ \citenamefont
  {Willcox}}]{farcasParametricNonintrusiveReducedorder2023}%
  \BibitemOpen
  \bibfield  {author} {\bibinfo {author} {\bibfnamefont {I.}~\bibnamefont
  {Farcas}}, \bibinfo {author} {\bibfnamefont {R.}~\bibnamefont {Gundevia}},
  \bibinfo {author} {\bibfnamefont {R.}~\bibnamefont {Munipalli}}, \ and\
  \bibinfo {author} {\bibfnamefont {K.~E.}\ \bibnamefont {Willcox}},\ }in\
  \href {\doibase 10.2514/6.2023-0172} {\emph {\bibinfo {booktitle} {{{AIAA
  SCITECH}} 2023 {{Forum}}}}}\ (\bibinfo  {publisher} {{American Institute of
  Aeronautics and Astronautics}},\ \bibinfo {year} {2023})\BibitemShut
  {NoStop}%
\bibitem [{\citenamefont {Qian}, \citenamefont {Farcas},\ and\ \citenamefont
  {Willcox}(2022)}]{qianReducedOperatorInference2022}%
  \BibitemOpen
  \bibfield  {author} {\bibinfo {author} {\bibfnamefont {E.}~\bibnamefont
  {Qian}}, \bibinfo {author} {\bibfnamefont {I.-G.}\ \bibnamefont {Farcas}}, \
  and\ \bibinfo {author} {\bibfnamefont {K.}~\bibnamefont {Willcox}},\
  }\href@noop {} {\bibfield  {journal} {\bibinfo  {journal} {SIAM Journal on
  Scientific Computing}\ }\textbf {\bibinfo {volume} {44}},\ \bibinfo {pages}
  {A1934} (\bibinfo {year} {2022})}\BibitemShut {NoStop}%
\bibitem [{\citenamefont {Swischuk}\ \emph {et~al.}(2020)\citenamefont
  {Swischuk}, \citenamefont {Kramer}, \citenamefont {Huang},\ and\
  \citenamefont {Willcox}}]{swischukLearningPhysicsbasedReducedorder2020}%
  \BibitemOpen
  \bibfield  {author} {\bibinfo {author} {\bibfnamefont {R.}~\bibnamefont
  {Swischuk}}, \bibinfo {author} {\bibfnamefont {B.}~\bibnamefont {Kramer}},
  \bibinfo {author} {\bibfnamefont {C.}~\bibnamefont {Huang}}, \ and\ \bibinfo
  {author} {\bibfnamefont {K.}~\bibnamefont {Willcox}},\ }\href {\doibase
  10.2514/1.J058943} {\bibfield  {journal} {\bibinfo  {journal} {AIAA Journal}\
  }\textbf {\bibinfo {volume} {58}},\ \bibinfo {pages} {2658} (\bibinfo {year}
  {2020})}\BibitemShut {NoStop}%
\bibitem [{\citenamefont {Issan}\ and\ \citenamefont
  {Kramer}(2023)}]{issanPredictingSolarWind2023}%
  \BibitemOpen
  \bibfield  {author} {\bibinfo {author} {\bibfnamefont {O.}~\bibnamefont
  {Issan}}\ and\ \bibinfo {author} {\bibfnamefont {B.}~\bibnamefont {Kramer}},\
  }\href {\doibase 10.1016/j.jcp.2022.111689} {\bibfield  {journal} {\bibinfo
  {journal} {Journal of Computational Physics}\ }\textbf {\bibinfo {volume}
  {473}},\ \bibinfo {pages} {111689} (\bibinfo {year} {2023})}\BibitemShut
  {NoStop}%
\bibitem [{\citenamefont {Khodabakhshi}\ and\ \citenamefont
  {Willcox}(2022)}]{khodabakhshiNonintrusiveDatadrivenModel2022}%
  \BibitemOpen
  \bibfield  {author} {\bibinfo {author} {\bibfnamefont {P.}~\bibnamefont
  {Khodabakhshi}}\ and\ \bibinfo {author} {\bibfnamefont {K.~E.}\ \bibnamefont
  {Willcox}},\ }\href {\doibase 10.1016/j.cma.2021.114296} {\bibfield
  {journal} {\bibinfo  {journal} {Computer Methods in Applied Mechanics and
  Engineering}\ }\textbf {\bibinfo {volume} {389}},\ \bibinfo {pages} {114296}
  (\bibinfo {year} {2022})}\BibitemShut {NoStop}%
\bibitem [{\citenamefont {Almeida}\ \emph {et~al.}(2022)\citenamefont
  {Almeida}, \citenamefont {Pires}, \citenamefont {Cid},\ and\ \citenamefont
  {Junior}}]{almeidaNonIntrusiveReducedModels2022}%
  \BibitemOpen
  \bibfield  {author} {\bibinfo {author} {\bibfnamefont {J.~L. d.~S.}\
  \bibnamefont {Almeida}}, \bibinfo {author} {\bibfnamefont {A.~C.}\
  \bibnamefont {Pires}}, \bibinfo {author} {\bibfnamefont {K.~F.~V.}\
  \bibnamefont {Cid}}, \ and\ \bibinfo {author} {\bibfnamefont {A.~C.~N.}\
  \bibnamefont {Junior}},\ }\href {\doibase 10.48550/arXiv.2206.01604}
  {\enquote {\bibinfo {title} {Non-{{Intrusive Reduced Models}} based on
  {{Operator Inference}} for {{Chaotic Systems}}},}\ } (\bibinfo {year}
  {2022}),\ \Eprint {http://arxiv.org/abs/2206.01604} {arXiv:2206.01604 [nlin]}
  \BibitemShut {NoStop}%
\bibitem [{\citenamefont {Farcas}, \citenamefont {Munipalli},\ and\
  \citenamefont {Willcox}(2022)}]{farcasFilteringNonintrusiveDatadriven2022}%
  \BibitemOpen
  \bibfield  {author} {\bibinfo {author} {\bibfnamefont {I.}~\bibnamefont
  {Farcas}}, \bibinfo {author} {\bibfnamefont {R.}~\bibnamefont {Munipalli}}, \
  and\ \bibinfo {author} {\bibfnamefont {K.~E.}\ \bibnamefont {Willcox}},\ }in\
  \href {\doibase 10.2514/6.2022-3487} {\emph {\bibinfo {booktitle} {{{AIAA
  AVIATION}} 2022 {{Forum}}}}},\ \bibinfo {series and number} {{{AIAA AVIATION
  Forum}}}\ (\bibinfo  {publisher} {{American Institute of Aeronautics and
  Astronautics}},\ \bibinfo {year} {2022})\BibitemShut {NoStop}%
\bibitem [{\citenamefont {Uy}, \citenamefont {Hartmann},\ and\ \citenamefont
  {Peherstorfer}(2023)}]{uyOperatorInferenceRoll2023}%
  \BibitemOpen
  \bibfield  {author} {\bibinfo {author} {\bibfnamefont {W.~I.~T.}\
  \bibnamefont {Uy}}, \bibinfo {author} {\bibfnamefont {D.}~\bibnamefont
  {Hartmann}}, \ and\ \bibinfo {author} {\bibfnamefont {B.}~\bibnamefont
  {Peherstorfer}},\ }\href {\doibase 10.1016/j.camwa.2023.06.012} {\bibfield
  {journal} {\bibinfo  {journal} {Computers \& Mathematics with Applications}\
  }\textbf {\bibinfo {volume} {145}},\ \bibinfo {pages} {224} (\bibinfo {year}
  {2023})}\BibitemShut {NoStop}%
\bibitem [{\citenamefont {Geelen}\ and\ \citenamefont
  {Willcox}(2022)}]{geelenLocalizedNonintrusiveReducedorder2022}%
  \BibitemOpen
  \bibfield  {author} {\bibinfo {author} {\bibfnamefont {R.}~\bibnamefont
  {Geelen}}\ and\ \bibinfo {author} {\bibfnamefont {K.}~\bibnamefont
  {Willcox}},\ }\href {\doibase 10.1098/rsta.2021.0206} {\bibfield  {journal}
  {\bibinfo  {journal} {Philosophical Transactions of the Royal Society A:
  Mathematical, Physical and Engineering Sciences}\ }\textbf {\bibinfo {volume}
  {380}},\ \bibinfo {pages} {20210206} (\bibinfo {year} {2022})}\BibitemShut
  {NoStop}%
\bibitem [{\citenamefont {Farcas}\ \emph
  {et~al.}(2024{\natexlab{b}})\citenamefont {Farcas}, \citenamefont {Gundevia},
  \citenamefont {Munipalli},\ and\ \citenamefont
  {Willcox}}]{farcasImprovingAccuracyScalability2023}%
  \BibitemOpen
  \bibfield  {author} {\bibinfo {author} {\bibfnamefont {I.-G.}\ \bibnamefont
  {Farcas}}, \bibinfo {author} {\bibfnamefont {R.~P.}\ \bibnamefont
  {Gundevia}}, \bibinfo {author} {\bibfnamefont {R.}~\bibnamefont {Munipalli}},
  \ and\ \bibinfo {author} {\bibfnamefont {K.~E.}\ \bibnamefont {Willcox}},\
  }\href {\doibase 10.2514/1.J063715} {\bibfield  {journal} {\bibinfo
  {journal} {AIAA Journal}\ ,\ \bibinfo {pages} {1}} (\bibinfo {year}
  {2024}{\natexlab{b}})}\BibitemShut {NoStop}%
\bibitem [{\citenamefont {Geelen}, \citenamefont {Wright},\ and\ \citenamefont
  {Willcox}(2023)}]{geelenOperatorInferenceNonintrusive2023}%
  \BibitemOpen
  \bibfield  {author} {\bibinfo {author} {\bibfnamefont {R.}~\bibnamefont
  {Geelen}}, \bibinfo {author} {\bibfnamefont {S.}~\bibnamefont {Wright}}, \
  and\ \bibinfo {author} {\bibfnamefont {K.}~\bibnamefont {Willcox}},\ }\href
  {\doibase 10.1016/j.cma.2022.115717} {\bibfield  {journal} {\bibinfo
  {journal} {Computer Methods in Applied Mechanics and Engineering}\ }\textbf
  {\bibinfo {volume} {403}},\ \bibinfo {pages} {115717} (\bibinfo {year}
  {2023})}\BibitemShut {NoStop}%
\bibitem [{\citenamefont {Hasegawa}\ and\ \citenamefont
  {Wakatani}(1983)}]{hasegawaPlasmaEdgeTurbulence1983}%
  \BibitemOpen
  \bibfield  {author} {\bibinfo {author} {\bibfnamefont {A.}~\bibnamefont
  {Hasegawa}}\ and\ \bibinfo {author} {\bibfnamefont {M.}~\bibnamefont
  {Wakatani}},\ }\href {\doibase 10.1103/PhysRevLett.50.682} {\bibfield
  {journal} {\bibinfo  {journal} {Physical Review Letters}\ }\textbf {\bibinfo
  {volume} {50}},\ \bibinfo {pages} {682} (\bibinfo {year} {1983})}\BibitemShut
  {NoStop}%
\bibitem [{\citenamefont {Camargo}, \citenamefont {Biskamp},\ and\
  \citenamefont {Scott}(1995)}]{camargoResistiveDriftWave1995}%
  \BibitemOpen
  \bibfield  {author} {\bibinfo {author} {\bibfnamefont {S.~J.}\ \bibnamefont
  {Camargo}}, \bibinfo {author} {\bibfnamefont {D.}~\bibnamefont {Biskamp}}, \
  and\ \bibinfo {author} {\bibfnamefont {B.~D.}\ \bibnamefont {Scott}},\ }\href
  {\doibase 10.1063/1.871116} {\bibfield  {journal} {\bibinfo  {journal}
  {Physics of Plasmas}\ }\textbf {\bibinfo {volume} {2}},\ \bibinfo {pages}
  {48} (\bibinfo {year} {1995})}\BibitemShut {NoStop}%
\bibitem [{\citenamefont {Manz}(2019)}]{manzMicroscopicPicturePlasma2019}%
  \BibitemOpen
  \bibfield  {author} {\bibinfo {author} {\bibfnamefont {P.}~\bibnamefont
  {Manz}},\ }\emph {\bibinfo {title} {The {{Microscopic Picture}} of {{Plasma
  Edge Turbulence}}}},\ \href@noop {} {\bibinfo {type} {Habilitation}},\
  \bibinfo  {school} {Technische Universit{\"a}t M{\"u}nchen} (\bibinfo {year}
  {2019})\BibitemShut {NoStop}%
\bibitem [{\citenamefont {Numata}, \citenamefont {Ball},\ and\ \citenamefont
  {Dewar}(2007)}]{numataBifurcationElectrostaticResistive2007}%
  \BibitemOpen
  \bibfield  {author} {\bibinfo {author} {\bibfnamefont {R.}~\bibnamefont
  {Numata}}, \bibinfo {author} {\bibfnamefont {R.}~\bibnamefont {Ball}}, \ and\
  \bibinfo {author} {\bibfnamefont {R.~L.}\ \bibnamefont {Dewar}},\ }\href
  {\doibase 10.1063/1.2796106} {\bibfield  {journal} {\bibinfo  {journal}
  {Physics of Plasmas}\ }\textbf {\bibinfo {volume} {14}},\ \bibinfo {pages}
  {102312} (\bibinfo {year} {2007})}\BibitemShut {NoStop}%
\bibitem [{\citenamefont {Sasaki}\ \emph {et~al.}(2020)\citenamefont {Sasaki},
  \citenamefont {Kobayashi}, \citenamefont {Dendy}, \citenamefont {Kawachi},
  \citenamefont {Arakawa},\ and\ \citenamefont
  {Inagaki}}]{sasakiEvaluationAbruptEnergy2020}%
  \BibitemOpen
  \bibfield  {author} {\bibinfo {author} {\bibfnamefont {M.}~\bibnamefont
  {Sasaki}}, \bibinfo {author} {\bibfnamefont {T.}~\bibnamefont {Kobayashi}},
  \bibinfo {author} {\bibfnamefont {R.~O.}\ \bibnamefont {Dendy}}, \bibinfo
  {author} {\bibfnamefont {Y.}~\bibnamefont {Kawachi}}, \bibinfo {author}
  {\bibfnamefont {H.}~\bibnamefont {Arakawa}}, \ and\ \bibinfo {author}
  {\bibfnamefont {S.}~\bibnamefont {Inagaki}},\ }\href {\doibase
  10.1088/1361-6587/abcb46} {\bibfield  {journal} {\bibinfo  {journal} {Plasma
  Physics and Controlled Fusion}\ }\textbf {\bibinfo {volume} {63}},\ \bibinfo
  {pages} {025004} (\bibinfo {year} {2020})}\BibitemShut {NoStop}%
\bibitem [{\citenamefont {Yatomi}, \citenamefont {Nakata},\ and\ \citenamefont
  {Sasaki}(2023)}]{yatomiDatadrivenModalAnalysis2023}%
  \BibitemOpen
  \bibfield  {author} {\bibinfo {author} {\bibfnamefont {G.}~\bibnamefont
  {Yatomi}}, \bibinfo {author} {\bibfnamefont {M.}~\bibnamefont {Nakata}}, \
  and\ \bibinfo {author} {\bibfnamefont {M.}~\bibnamefont {Sasaki}},\ }\href
  {\doibase 10.1088/1361-6587/ace993} {\bibfield  {journal} {\bibinfo
  {journal} {Plasma Physics and Controlled Fusion}\ }\textbf {\bibinfo {volume}
  {65}},\ \bibinfo {pages} {095014} (\bibinfo {year} {2023})}\BibitemShut
  {NoStop}%
\bibitem [{\citenamefont {Berkooz}, \citenamefont {Holmes},\ and\ \citenamefont
  {Lumley}(1993)}]{berkoozProperOrthogonalDecomposition1993}%
  \BibitemOpen
  \bibfield  {author} {\bibinfo {author} {\bibfnamefont {G.}~\bibnamefont
  {Berkooz}}, \bibinfo {author} {\bibfnamefont {P.}~\bibnamefont {Holmes}}, \
  and\ \bibinfo {author} {\bibfnamefont {J.~L.}\ \bibnamefont {Lumley}},\
  }\href {\doibase 10.1146/annurev.fl.25.010193.002543} {\bibfield  {journal}
  {\bibinfo  {journal} {Annual Review of Fluid Mechanics}\ }\textbf {\bibinfo
  {volume} {25}},\ \bibinfo {pages} {539} (\bibinfo {year} {1993})}\BibitemShut
  {NoStop}%
\bibitem [{\citenamefont {Goumiri}\ \emph {et~al.}(2013)\citenamefont
  {Goumiri}, \citenamefont {Rowley}, \citenamefont {Ma}, \citenamefont {Gates},
  \citenamefont {Krommes},\ and\ \citenamefont
  {Parker}}]{goumiriReducedorderModelBased2013}%
  \BibitemOpen
  \bibfield  {author} {\bibinfo {author} {\bibfnamefont {I.~R.}\ \bibnamefont
  {Goumiri}}, \bibinfo {author} {\bibfnamefont {C.~W.}\ \bibnamefont {Rowley}},
  \bibinfo {author} {\bibfnamefont {Z.}~\bibnamefont {Ma}}, \bibinfo {author}
  {\bibfnamefont {D.~A.}\ \bibnamefont {Gates}}, \bibinfo {author}
  {\bibfnamefont {J.~A.}\ \bibnamefont {Krommes}}, \ and\ \bibinfo {author}
  {\bibfnamefont {J.~B.}\ \bibnamefont {Parker}},\ }\href {\doibase
  10.1063/1.4796190} {\bibfield  {journal} {\bibinfo  {journal} {Physics of
  Plasmas}\ }\textbf {\bibinfo {volume} {20}},\ \bibinfo {pages} {042501}
  (\bibinfo {year} {2013})}\BibitemShut {NoStop}%
\bibitem [{\citenamefont {Castagna}\ \emph {et~al.}(2024)\citenamefont
  {Castagna}, \citenamefont {Schiavello}, \citenamefont {Zanisi},\ and\
  \citenamefont {Williams}}]{castagnaStyleGANAIDeconvolution2024}%
  \BibitemOpen
  \bibfield  {author} {\bibinfo {author} {\bibfnamefont {J.}~\bibnamefont
  {Castagna}}, \bibinfo {author} {\bibfnamefont {F.}~\bibnamefont
  {Schiavello}}, \bibinfo {author} {\bibfnamefont {L.}~\bibnamefont {Zanisi}},
  \ and\ \bibinfo {author} {\bibfnamefont {J.}~\bibnamefont {Williams}},\
  }\href {\doibase 10.1063/5.0189945} {\bibfield  {journal} {\bibinfo
  {journal} {Physics of Plasmas}\ }\textbf {\bibinfo {volume} {31}},\ \bibinfo
  {pages} {033902} (\bibinfo {year} {2024})}\BibitemShut {NoStop}%
\bibitem [{\citenamefont {Greif}, \citenamefont {Jenko},\ and\ \citenamefont
  {Thuerey}(2023)}]{greifPhysicsPreservingAIAcceleratedSimulations2023}%
  \BibitemOpen
  \bibfield  {author} {\bibinfo {author} {\bibfnamefont {R.}~\bibnamefont
  {Greif}}, \bibinfo {author} {\bibfnamefont {F.}~\bibnamefont {Jenko}}, \ and\
  \bibinfo {author} {\bibfnamefont {N.}~\bibnamefont {Thuerey}},\ }\href
  {\doibase 10.48550/arXiv.2309.16400} {\enquote {\bibinfo {title}
  {Physics-{{Preserving AI-Accelerated Simulations}} of {{Plasma
  Turbulence}}},}\ } (\bibinfo {year} {2023}),\ \Eprint
  {http://arxiv.org/abs/2309.16400} {arXiv:2309.16400 [physics]} \BibitemShut
  {NoStop}%
\bibitem [{\citenamefont {Clavier}\ \emph {et~al.}(2024)\citenamefont
  {Clavier}, \citenamefont {Zarzoso}, \citenamefont {{del-Castillo-Negrete}},\
  and\ \citenamefont {Frenord}}]{clavierGenerativeMachineLearning2024}%
  \BibitemOpen
  \bibfield  {author} {\bibinfo {author} {\bibfnamefont {B.}~\bibnamefont
  {Clavier}}, \bibinfo {author} {\bibfnamefont {D.}~\bibnamefont {Zarzoso}},
  \bibinfo {author} {\bibfnamefont {D.}~\bibnamefont {{del-Castillo-Negrete}}},
  \ and\ \bibinfo {author} {\bibfnamefont {E.}~\bibnamefont {Frenord}},\ }\href
  {\doibase 10.48550/arXiv.2405.13232} {\enquote {\bibinfo {title} {A
  generative machine learning surrogate model of plasma turbulence},}\ }
  (\bibinfo {year} {2024}),\ \Eprint {http://arxiv.org/abs/2405.13232}
  {arXiv:2405.13232 [physics]} \BibitemShut {NoStop}%
\bibitem [{\citenamefont {Heinonen}\ and\ \citenamefont
  {Diamond}(2020)}]{heinonenTurbulenceModelReduction2020}%
  \BibitemOpen
  \bibfield  {author} {\bibinfo {author} {\bibfnamefont {R.~A.}\ \bibnamefont
  {Heinonen}}\ and\ \bibinfo {author} {\bibfnamefont {P.~H.}\ \bibnamefont
  {Diamond}},\ }\href {\doibase 10.1103/PhysRevE.101.061201} {\bibfield
  {journal} {\bibinfo  {journal} {Physical Review E}\ }\textbf {\bibinfo
  {volume} {101}},\ \bibinfo {pages} {061201} (\bibinfo {year}
  {2020})}\BibitemShut {NoStop}%
\bibitem [{\citenamefont {Lin}\ \emph {et~al.}(2024)\citenamefont {Lin},
  \citenamefont {{Maurel-Oujia}}, \citenamefont {Kadoch}, \citenamefont {Krah},
  \citenamefont {Saura}, \citenamefont {Benkadda},\ and\ \citenamefont
  {Schneider}}]{linSynthesizingImpurityClustering2024}%
  \BibitemOpen
  \bibfield  {author} {\bibinfo {author} {\bibfnamefont {Z.}~\bibnamefont
  {Lin}}, \bibinfo {author} {\bibfnamefont {T.}~\bibnamefont {{Maurel-Oujia}}},
  \bibinfo {author} {\bibfnamefont {B.}~\bibnamefont {Kadoch}}, \bibinfo
  {author} {\bibfnamefont {P.}~\bibnamefont {Krah}}, \bibinfo {author}
  {\bibfnamefont {N.}~\bibnamefont {Saura}}, \bibinfo {author} {\bibfnamefont
  {S.}~\bibnamefont {Benkadda}}, \ and\ \bibinfo {author} {\bibfnamefont
  {K.}~\bibnamefont {Schneider}},\ }\href {\doibase 10.1063/5.0178085}
  {\bibfield  {journal} {\bibinfo  {journal} {Physics of Plasmas}\ }\textbf
  {\bibinfo {volume} {31}},\ \bibinfo {pages} {032505} (\bibinfo {year}
  {2024})}\BibitemShut {NoStop}%
\bibitem [{\citenamefont {Faraji}\ \emph
  {et~al.}(2023{\natexlab{a}})\citenamefont {Faraji}, \citenamefont {Reza},
  \citenamefont {Knoll},\ and\ \citenamefont
  {Kutz}}]{farajiDynamicModeDecomposition2023}%
  \BibitemOpen
  \bibfield  {author} {\bibinfo {author} {\bibfnamefont {F.}~\bibnamefont
  {Faraji}}, \bibinfo {author} {\bibfnamefont {M.}~\bibnamefont {Reza}},
  \bibinfo {author} {\bibfnamefont {A.}~\bibnamefont {Knoll}}, \ and\ \bibinfo
  {author} {\bibfnamefont {J.~N.}\ \bibnamefont {Kutz}},\ }\href {\doibase
  10.1088/1361-6463/ad0910} {\bibfield  {journal} {\bibinfo  {journal} {Journal
  of Physics D: Applied Physics}\ }\textbf {\bibinfo {volume} {57}},\ \bibinfo
  {pages} {065201} (\bibinfo {year} {2023}{\natexlab{a}})}\BibitemShut
  {NoStop}%
\bibitem [{\citenamefont {Futatani}, \citenamefont {Benkadda},\ and\
  \citenamefont
  {{del-Castillo-Negrete}}(2009)}]{futataniSpatiotemporalMultiscalingAnalysis2009}%
  \BibitemOpen
  \bibfield  {author} {\bibinfo {author} {\bibfnamefont {S.}~\bibnamefont
  {Futatani}}, \bibinfo {author} {\bibfnamefont {S.}~\bibnamefont {Benkadda}},
  \ and\ \bibinfo {author} {\bibfnamefont {D.}~\bibnamefont
  {{del-Castillo-Negrete}}},\ }\href {\doibase 10.1063/1.3095865} {\bibfield
  {journal} {\bibinfo  {journal} {Physics of Plasmas}\ }\textbf {\bibinfo
  {volume} {16}},\ \bibinfo {pages} {042506} (\bibinfo {year}
  {2009})}\BibitemShut {NoStop}%
\bibitem [{\citenamefont {Kusaba}, \citenamefont {Kuboyama},\ and\
  \citenamefont {Inagaki}(2020)}]{kusabaSparsityPromotingDynamicMode2020}%
  \BibitemOpen
  \bibfield  {author} {\bibinfo {author} {\bibfnamefont {A.}~\bibnamefont
  {Kusaba}}, \bibinfo {author} {\bibfnamefont {T.}~\bibnamefont {Kuboyama}}, \
  and\ \bibinfo {author} {\bibfnamefont {S.}~\bibnamefont {Inagaki}},\ }\href
  {\doibase 10.1585/pfr.15.1301001} {\bibfield  {journal} {\bibinfo  {journal}
  {Plasma and Fusion Research}\ }\textbf {\bibinfo {volume} {15}},\ \bibinfo
  {pages} {1301001} (\bibinfo {year} {2020})}\BibitemShut {NoStop}%
\bibitem [{\citenamefont {Taylor}\ \emph {et~al.}(2018)\citenamefont {Taylor},
  \citenamefont {Kutz}, \citenamefont {Morgan},\ and\ \citenamefont
  {Nelson}}]{taylorDynamicModeDecomposition2018}%
  \BibitemOpen
  \bibfield  {author} {\bibinfo {author} {\bibfnamefont {R.}~\bibnamefont
  {Taylor}}, \bibinfo {author} {\bibfnamefont {J.~N.}\ \bibnamefont {Kutz}},
  \bibinfo {author} {\bibfnamefont {K.}~\bibnamefont {Morgan}}, \ and\ \bibinfo
  {author} {\bibfnamefont {B.}~\bibnamefont {Nelson}},\ }\href {\doibase
  10.1063/1.5027419} {\bibfield  {journal} {\bibinfo  {journal} {Review of
  Scientific Instruments}\ }\textbf {\bibinfo {volume} {89}},\ \bibinfo {pages}
  {053501} (\bibinfo {year} {2018})},\ \Eprint
  {http://arxiv.org/abs/1702.06871} {arXiv:1702.06871} \BibitemShut {NoStop}%
\bibitem [{\citenamefont {Faraji}\ \emph
  {et~al.}(2023{\natexlab{b}})\citenamefont {Faraji}, \citenamefont {Reza},
  \citenamefont {Knoll},\ and\ \citenamefont
  {Kutz}}]{farajiDynamicModeDecomposition2023II}%
  \BibitemOpen
  \bibfield  {author} {\bibinfo {author} {\bibfnamefont {F.}~\bibnamefont
  {Faraji}}, \bibinfo {author} {\bibfnamefont {M.}~\bibnamefont {Reza}},
  \bibinfo {author} {\bibfnamefont {A.}~\bibnamefont {Knoll}}, \ and\ \bibinfo
  {author} {\bibfnamefont {J.~N.}\ \bibnamefont {Kutz}},\ }\href {\doibase
  10.1088/1361-6463/ad0911} {\bibfield  {journal} {\bibinfo  {journal} {Journal
  of Physics D: Applied Physics}\ }\textbf {\bibinfo {volume} {57}},\ \bibinfo
  {pages} {065202} (\bibinfo {year} {2023}{\natexlab{b}})}\BibitemShut
  {NoStop}%
\bibitem [{\citenamefont {Kaptanoglu}\ \emph {et~al.}(2020)\citenamefont
  {Kaptanoglu}, \citenamefont {Morgan}, \citenamefont {Hansen},\ and\
  \citenamefont {Brunton}}]{kaptanogluCharacterizingMagnetizedPlasmas2020}%
  \BibitemOpen
  \bibfield  {author} {\bibinfo {author} {\bibfnamefont {A.~A.}\ \bibnamefont
  {Kaptanoglu}}, \bibinfo {author} {\bibfnamefont {K.~D.}\ \bibnamefont
  {Morgan}}, \bibinfo {author} {\bibfnamefont {C.~J.}\ \bibnamefont {Hansen}},
  \ and\ \bibinfo {author} {\bibfnamefont {S.~L.}\ \bibnamefont {Brunton}},\
  }\href {\doibase 10.1063/1.5138932} {\bibfield  {journal} {\bibinfo
  {journal} {Physics of Plasmas}\ }\textbf {\bibinfo {volume} {27}},\ \bibinfo
  {pages} {032108} (\bibinfo {year} {2020})}\BibitemShut {NoStop}%
\bibitem [{\citenamefont {Jajima}\ \emph {et~al.}(2023)\citenamefont {Jajima},
  \citenamefont {Sasaki}, \citenamefont {Ishikawa}, \citenamefont {Nakata},
  \citenamefont {Kobayashi}, \citenamefont {Kawachi},\ and\ \citenamefont
  {Arakawa}}]{jajimaEstimation2DProfile2023}%
  \BibitemOpen
  \bibfield  {author} {\bibinfo {author} {\bibfnamefont {Y.}~\bibnamefont
  {Jajima}}, \bibinfo {author} {\bibfnamefont {M.}~\bibnamefont {Sasaki}},
  \bibinfo {author} {\bibfnamefont {R.~T.}\ \bibnamefont {Ishikawa}}, \bibinfo
  {author} {\bibfnamefont {M.}~\bibnamefont {Nakata}}, \bibinfo {author}
  {\bibfnamefont {T.}~\bibnamefont {Kobayashi}}, \bibinfo {author}
  {\bibfnamefont {Y.}~\bibnamefont {Kawachi}}, \ and\ \bibinfo {author}
  {\bibfnamefont {H.}~\bibnamefont {Arakawa}},\ }\href {\doibase
  10.1088/1361-6587/acff7f} {\bibfield  {journal} {\bibinfo  {journal} {Plasma
  Physics and Controlled Fusion}\ }\textbf {\bibinfo {volume} {65}},\ \bibinfo
  {pages} {125003} (\bibinfo {year} {2023})}\BibitemShut {NoStop}%
\bibitem [{\citenamefont {Gopakumar}\ \emph {et~al.}(2023)\citenamefont
  {Gopakumar}, \citenamefont {Pamela}, \citenamefont {Zanisi}, \citenamefont
  {Li}, \citenamefont {Anandkumar},\ and\ \citenamefont
  {Team}}]{gopakumarFourierNeuralOperator2023}%
  \BibitemOpen
  \bibfield  {author} {\bibinfo {author} {\bibfnamefont {V.}~\bibnamefont
  {Gopakumar}}, \bibinfo {author} {\bibfnamefont {S.}~\bibnamefont {Pamela}},
  \bibinfo {author} {\bibfnamefont {L.}~\bibnamefont {Zanisi}}, \bibinfo
  {author} {\bibfnamefont {Z.}~\bibnamefont {Li}}, \bibinfo {author}
  {\bibfnamefont {A.}~\bibnamefont {Anandkumar}}, \ and\ \bibinfo {author}
  {\bibfnamefont {{\relax MAST}.}~\bibnamefont {Team}},\ }\href@noop {}
  {\enquote {\bibinfo {title} {{Fourier Neural Operator for Plasma
  Modelling}},}\ } (\bibinfo {year} {2023}),\ \Eprint
  {http://arxiv.org/abs/2302.06542} {arXiv:2302.06542 [physics]} \BibitemShut
  {NoStop}%
\bibitem [{\citenamefont {Ghazijahani}\ and\ \citenamefont
  {Cierpka}(2024)}]{ghazijahaniPredictionTurbulentFlow2024}%
  \BibitemOpen
  \bibfield  {author} {\bibinfo {author} {\bibfnamefont {M.~S.}\ \bibnamefont
  {Ghazijahani}}\ and\ \bibinfo {author} {\bibfnamefont {C.}~\bibnamefont
  {Cierpka}},\ }\href {\doibase 10.1088/2632-2153/ad5414} {\bibfield  {journal}
  {\bibinfo  {journal} {Machine Learning: Science and Technology}\ }\textbf
  {\bibinfo {volume} {5}},\ \bibinfo {pages} {035005} (\bibinfo {year}
  {2024})}\BibitemShut {NoStop}%
\bibitem [{\citenamefont {Wakatani}\ and\ \citenamefont
  {Hasegawa}(1984)}]{wakataniCollisionalDriftWave1984}%
  \BibitemOpen
  \bibfield  {author} {\bibinfo {author} {\bibfnamefont {M.}~\bibnamefont
  {Wakatani}}\ and\ \bibinfo {author} {\bibfnamefont {A.}~\bibnamefont
  {Hasegawa}},\ }\href {\doibase 10.1063/1.864660} {\bibfield  {journal}
  {\bibinfo  {journal} {The Physics of Fluids}\ }\textbf {\bibinfo {volume}
  {27}},\ \bibinfo {pages} {611} (\bibinfo {year} {1984})}\BibitemShut
  {NoStop}%
\bibitem [{\citenamefont {Hasegawa}\ and\ \citenamefont
  {Mima}(1978)}]{hasegawaPseudoThreeDimensional1978}%
  \BibitemOpen
  \bibfield  {author} {\bibinfo {author} {\bibfnamefont {A.}~\bibnamefont
  {Hasegawa}}\ and\ \bibinfo {author} {\bibfnamefont {K.}~\bibnamefont
  {Mima}},\ }\href {\doibase 10.1063/1.862083} {\bibfield  {journal} {\bibinfo
  {journal} {Physics of Fluids}\ }\textbf {\bibinfo {volume} {21}},\ \bibinfo
  {pages} {87} (\bibinfo {year} {1978})}\BibitemShut {NoStop}%
\bibitem [{\citenamefont {Kim}\ and\ \citenamefont
  {Terry}(2013)}]{kimNumericalInvestigationFrequency2013}%
  \BibitemOpen
  \bibfield  {author} {\bibinfo {author} {\bibfnamefont {J.}~\bibnamefont
  {Kim}}\ and\ \bibinfo {author} {\bibfnamefont {P.~W.}\ \bibnamefont
  {Terry}},\ }\href {\doibase 10.1063/1.4822335} {\bibfield  {journal}
  {\bibinfo  {journal} {Physics of Plasmas}\ }\textbf {\bibinfo {volume}
  {20}},\ \bibinfo {pages} {102303} (\bibinfo {year} {2013})}\BibitemShut
  {NoStop}%
\bibitem [{\citenamefont {Greif}(2023)}]{greifHW2DReferenceImplementation2023}%
  \BibitemOpen
  \bibfield  {author} {\bibinfo {author} {\bibfnamefont {R.}~\bibnamefont
  {Greif}},\ }\href {\doibase 10.21105/joss.05959} {\bibfield  {journal}
  {\bibinfo  {journal} {Journal of Open Source Software}\ }\textbf {\bibinfo
  {volume} {8}},\ \bibinfo {pages} {5959} (\bibinfo {year} {2023})}\BibitemShut
  {NoStop}%
\bibitem [{\citenamefont
  {Arakawa}(1997)}]{arakawaComputationalDesignLongTerm1997}%
  \BibitemOpen
  \bibfield  {author} {\bibinfo {author} {\bibfnamefont {A.}~\bibnamefont
  {Arakawa}},\ }\href@noop {} {\bibfield  {journal} {\bibinfo  {journal}
  {Journal of computational physics}\ }\textbf {\bibinfo {volume} {135}},\
  \bibinfo {pages} {103} (\bibinfo {year} {1997})}\BibitemShut {NoStop}%
\bibitem [{\citenamefont {Runge}(1895)}]{rungeUeberNumerischeAufloesung1895}%
  \BibitemOpen
  \bibfield  {author} {\bibinfo {author} {\bibfnamefont {C.}~\bibnamefont
  {Runge}},\ }\href {\doibase 10.1007/BF01446807} {\bibfield  {journal}
  {\bibinfo  {journal} {Mathematische Annalen}\ }\textbf {\bibinfo {volume}
  {46}},\ \bibinfo {pages} {167} (\bibinfo {year} {1895})}\BibitemShut
  {NoStop}%
\bibitem [{\citenamefont {Golub}\ and\ \citenamefont
  {Loan}(2013)}]{golubMatrixComputations2013}%
  \BibitemOpen
  \bibfield  {author} {\bibinfo {author} {\bibfnamefont {G.~H.}\ \bibnamefont
  {Golub}}\ and\ \bibinfo {author} {\bibfnamefont {C.~F.~V.}\ \bibnamefont
  {Loan}},\ }\href@noop {} {\emph {\bibinfo {title} {{Matrix Computations}}}}\
  (\bibinfo  {publisher} {The Johns Hopkins University Press},\ \bibinfo {year}
  {2013})\BibitemShut {NoStop}%
\bibitem [{\citenamefont {McQuarrie}, \citenamefont {Huang},\ and\
  \citenamefont {Willcox}(2021)}]{mcquarrieDatadrivenReducedorderModels2021}%
  \BibitemOpen
  \bibfield  {author} {\bibinfo {author} {\bibfnamefont {S.~A.}\ \bibnamefont
  {McQuarrie}}, \bibinfo {author} {\bibfnamefont {C.}~\bibnamefont {Huang}}, \
  and\ \bibinfo {author} {\bibfnamefont {K.~E.}\ \bibnamefont {Willcox}},\
  }\href {\doibase 10.1080/03036758.2020.1863237} {\bibfield  {journal}
  {\bibinfo  {journal} {Journal of the Royal Society of New Zealand}\ }\textbf
  {\bibinfo {volume} {51}},\ \bibinfo {pages} {194} (\bibinfo {year}
  {2021})}\BibitemShut {NoStop}%
\bibitem [{\citenamefont {Li}\ \emph {et~al.}(2023)\citenamefont {Li},
  \citenamefont {Buzzicotti}, \citenamefont {Biferale}, \citenamefont
  {Bonaccorso}, \citenamefont {Chen},\ and\ \citenamefont
  {Wan}}]{liMultiscaleReconstructionTurbulent2023}%
  \BibitemOpen
  \bibfield  {author} {\bibinfo {author} {\bibfnamefont {T.}~\bibnamefont
  {Li}}, \bibinfo {author} {\bibfnamefont {M.}~\bibnamefont {Buzzicotti}},
  \bibinfo {author} {\bibfnamefont {L.}~\bibnamefont {Biferale}}, \bibinfo
  {author} {\bibfnamefont {F.}~\bibnamefont {Bonaccorso}}, \bibinfo {author}
  {\bibfnamefont {S.}~\bibnamefont {Chen}}, \ and\ \bibinfo {author}
  {\bibfnamefont {M.}~\bibnamefont {Wan}},\ }\href {\doibase
  10.1017/jfm.2023.573} {\bibfield  {journal} {\bibinfo  {journal} {Journal of
  Fluid Mechanics}\ }\textbf {\bibinfo {volume} {971}},\ \bibinfo {pages} {A3}
  (\bibinfo {year} {2023})}\BibitemShut {NoStop}%
\bibitem [{\citenamefont {Cazelles}, \citenamefont {Robert},\ and\
  \citenamefont
  {Tobar}(2021)}]{cazellesWassersteinFourierDistanceStationary2021}%
  \BibitemOpen
  \bibfield  {author} {\bibinfo {author} {\bibfnamefont {E.}~\bibnamefont
  {Cazelles}}, \bibinfo {author} {\bibfnamefont {A.}~\bibnamefont {Robert}}, \
  and\ \bibinfo {author} {\bibfnamefont {F.}~\bibnamefont {Tobar}},\ }\href
  {\doibase 10.1109/TSP.2020.3046227} {\bibfield  {journal} {\bibinfo
  {journal} {IEEE Transactions on Signal Processing}\ }\textbf {\bibinfo
  {volume} {69}},\ \bibinfo {pages} {709} (\bibinfo {year} {2021})}\BibitemShut
  {NoStop}%
\bibitem [{\citenamefont {Bruna}, \citenamefont {Peherstorfer},\ and\
  \citenamefont {{Vanden-Eijnden}}(2024)}]{brunaNeuralGalerkinSchemes2024}%
  \BibitemOpen
  \bibfield  {author} {\bibinfo {author} {\bibfnamefont {J.}~\bibnamefont
  {Bruna}}, \bibinfo {author} {\bibfnamefont {B.}~\bibnamefont {Peherstorfer}},
  \ and\ \bibinfo {author} {\bibfnamefont {E.}~\bibnamefont
  {{Vanden-Eijnden}}},\ }\href {\doibase 10.1016/j.jcp.2023.112588} {\bibfield
  {journal} {\bibinfo  {journal} {Journal of Computational Physics}\ }\textbf
  {\bibinfo {volume} {496}},\ \bibinfo {pages} {112588} (\bibinfo {year}
  {2024})}\BibitemShut {NoStop}%
\bibitem [{\citenamefont
  {Peherstorfer}(2020)}]{peherstorferModelReductionTransportDominated2020}%
  \BibitemOpen
  \bibfield  {author} {\bibinfo {author} {\bibfnamefont {B.}~\bibnamefont
  {Peherstorfer}},\ }\href {\doibase 10.1137/19M1257275} {\bibfield  {journal}
  {\bibinfo  {journal} {SIAM Journal on Scientific Computing}\ }\textbf
  {\bibinfo {volume} {42}},\ \bibinfo {pages} {A2803} (\bibinfo {year}
  {2020})}\BibitemShut {NoStop}%
\bibitem [{\citenamefont {Harris}\ \emph {et~al.}(2020)\citenamefont {Harris},
  \citenamefont {Millman}, \citenamefont {{van der Walt}}, \citenamefont
  {Gommers}, \citenamefont {Virtanen}, \citenamefont {Cournapeau},
  \citenamefont {Wieser}, \citenamefont {Taylor}, \citenamefont {Berg},
  \citenamefont {Smith}, \citenamefont {Kern}, \citenamefont {Picus},
  \citenamefont {Hoyer}, \citenamefont {{van Kerkwijk}}, \citenamefont {Brett},
  \citenamefont {Haldane}, \citenamefont {{del R{\'i}o}}, \citenamefont
  {Wiebe}, \citenamefont {Peterson}, \citenamefont {{G{\'e}rard-Marchant}},
  \citenamefont {Sheppard}, \citenamefont {Reddy}, \citenamefont {Weckesser},
  \citenamefont {Abbasi}, \citenamefont {Gohlke},\ and\ \citenamefont
  {Oliphant}}]{harrisArrayProgrammingNumPy2020}%
  \BibitemOpen
  \bibfield  {author} {\bibinfo {author} {\bibfnamefont {C.~R.}\ \bibnamefont
  {Harris}}, \bibinfo {author} {\bibfnamefont {K.~J.}\ \bibnamefont {Millman}},
  \bibinfo {author} {\bibfnamefont {S.~J.}\ \bibnamefont {{van der Walt}}},
  \bibinfo {author} {\bibfnamefont {R.}~\bibnamefont {Gommers}}, \bibinfo
  {author} {\bibfnamefont {P.}~\bibnamefont {Virtanen}}, \bibinfo {author}
  {\bibfnamefont {D.}~\bibnamefont {Cournapeau}}, \bibinfo {author}
  {\bibfnamefont {E.}~\bibnamefont {Wieser}}, \bibinfo {author} {\bibfnamefont
  {J.}~\bibnamefont {Taylor}}, \bibinfo {author} {\bibfnamefont
  {S.}~\bibnamefont {Berg}}, \bibinfo {author} {\bibfnamefont {N.~J.}\
  \bibnamefont {Smith}}, \bibinfo {author} {\bibfnamefont {R.}~\bibnamefont
  {Kern}}, \bibinfo {author} {\bibfnamefont {M.}~\bibnamefont {Picus}},
  \bibinfo {author} {\bibfnamefont {S.}~\bibnamefont {Hoyer}}, \bibinfo
  {author} {\bibfnamefont {M.~H.}\ \bibnamefont {{van Kerkwijk}}}, \bibinfo
  {author} {\bibfnamefont {M.}~\bibnamefont {Brett}}, \bibinfo {author}
  {\bibfnamefont {A.}~\bibnamefont {Haldane}}, \bibinfo {author} {\bibfnamefont
  {J.~F.}\ \bibnamefont {{del R{\'i}o}}}, \bibinfo {author} {\bibfnamefont
  {M.}~\bibnamefont {Wiebe}}, \bibinfo {author} {\bibfnamefont
  {P.}~\bibnamefont {Peterson}}, \bibinfo {author} {\bibfnamefont
  {P.}~\bibnamefont {{G{\'e}rard-Marchant}}}, \bibinfo {author} {\bibfnamefont
  {K.}~\bibnamefont {Sheppard}}, \bibinfo {author} {\bibfnamefont
  {T.}~\bibnamefont {Reddy}}, \bibinfo {author} {\bibfnamefont
  {W.}~\bibnamefont {Weckesser}}, \bibinfo {author} {\bibfnamefont
  {H.}~\bibnamefont {Abbasi}}, \bibinfo {author} {\bibfnamefont
  {C.}~\bibnamefont {Gohlke}}, \ and\ \bibinfo {author} {\bibfnamefont {T.~E.}\
  \bibnamefont {Oliphant}},\ }\href {\doibase 10.1038/s41586-020-2649-2}
  {\bibfield  {journal} {\bibinfo  {journal} {Nature}\ }\textbf {\bibinfo
  {volume} {585}},\ \bibinfo {pages} {357} (\bibinfo {year}
  {2020})}\BibitemShut {NoStop}%
\bibitem [{\citenamefont {Virtanen}\ \emph {et~al.}(2020)\citenamefont
  {Virtanen}, \citenamefont {Gommers}, \citenamefont {Oliphant}, \citenamefont
  {Haberland}, \citenamefont {Reddy}, \citenamefont {Cournapeau}, \citenamefont
  {Burovski}, \citenamefont {Peterson}, \citenamefont {Weckesser},
  \citenamefont {Bright}, \citenamefont {{van der Walt}}, \citenamefont
  {Brett}, \citenamefont {Wilson}, \citenamefont {Millman}, \citenamefont
  {Mayorov}, \citenamefont {Nelson}, \citenamefont {Jones}, \citenamefont
  {Kern}, \citenamefont {Larson}, \citenamefont {Carey}, \citenamefont {Polat},
  \citenamefont {Feng}, \citenamefont {Moore}, \citenamefont {VanderPlas},
  \citenamefont {Laxalde}, \citenamefont {Perktold}, \citenamefont {Cimrman},
  \citenamefont {Henriksen}, \citenamefont {Quintero}, \citenamefont {Harris},
  \citenamefont {Archibald}, \citenamefont {Ribeiro}, \citenamefont
  {Pedregosa},\ and\ \citenamefont {{van
  Mulbregt}}}]{virtanenSciPyFundamentalAlgorithms2020}%
  \BibitemOpen
  \bibfield  {author} {\bibinfo {author} {\bibfnamefont {P.}~\bibnamefont
  {Virtanen}}, \bibinfo {author} {\bibfnamefont {R.}~\bibnamefont {Gommers}},
  \bibinfo {author} {\bibfnamefont {T.~E.}\ \bibnamefont {Oliphant}}, \bibinfo
  {author} {\bibfnamefont {M.}~\bibnamefont {Haberland}}, \bibinfo {author}
  {\bibfnamefont {T.}~\bibnamefont {Reddy}}, \bibinfo {author} {\bibfnamefont
  {D.}~\bibnamefont {Cournapeau}}, \bibinfo {author} {\bibfnamefont
  {E.}~\bibnamefont {Burovski}}, \bibinfo {author} {\bibfnamefont
  {P.}~\bibnamefont {Peterson}}, \bibinfo {author} {\bibfnamefont
  {W.}~\bibnamefont {Weckesser}}, \bibinfo {author} {\bibfnamefont
  {J.}~\bibnamefont {Bright}}, \bibinfo {author} {\bibfnamefont {S.~J.}\
  \bibnamefont {{van der Walt}}}, \bibinfo {author} {\bibfnamefont
  {M.}~\bibnamefont {Brett}}, \bibinfo {author} {\bibfnamefont
  {J.}~\bibnamefont {Wilson}}, \bibinfo {author} {\bibfnamefont {K.~J.}\
  \bibnamefont {Millman}}, \bibinfo {author} {\bibfnamefont {N.}~\bibnamefont
  {Mayorov}}, \bibinfo {author} {\bibfnamefont {A.~R.~J.}\ \bibnamefont
  {Nelson}}, \bibinfo {author} {\bibfnamefont {E.}~\bibnamefont {Jones}},
  \bibinfo {author} {\bibfnamefont {R.}~\bibnamefont {Kern}}, \bibinfo {author}
  {\bibfnamefont {E.}~\bibnamefont {Larson}}, \bibinfo {author} {\bibfnamefont
  {C.~J.}\ \bibnamefont {Carey}}, \bibinfo {author} {\bibfnamefont
  {{\.I}.}~\bibnamefont {Polat}}, \bibinfo {author} {\bibfnamefont
  {Y.}~\bibnamefont {Feng}}, \bibinfo {author} {\bibfnamefont {E.~W.}\
  \bibnamefont {Moore}}, \bibinfo {author} {\bibfnamefont {J.}~\bibnamefont
  {VanderPlas}}, \bibinfo {author} {\bibfnamefont {D.}~\bibnamefont {Laxalde}},
  \bibinfo {author} {\bibfnamefont {J.}~\bibnamefont {Perktold}}, \bibinfo
  {author} {\bibfnamefont {R.}~\bibnamefont {Cimrman}}, \bibinfo {author}
  {\bibfnamefont {I.}~\bibnamefont {Henriksen}}, \bibinfo {author}
  {\bibfnamefont {E.~A.}\ \bibnamefont {Quintero}}, \bibinfo {author}
  {\bibfnamefont {C.~R.}\ \bibnamefont {Harris}}, \bibinfo {author}
  {\bibfnamefont {A.~M.}\ \bibnamefont {Archibald}}, \bibinfo {author}
  {\bibfnamefont {A.~H.}\ \bibnamefont {Ribeiro}}, \bibinfo {author}
  {\bibfnamefont {F.}~\bibnamefont {Pedregosa}}, \ and\ \bibinfo {author}
  {\bibfnamefont {P.}~\bibnamefont {{van Mulbregt}}},\ }\href {\doibase
  10.1038/s41592-019-0686-2} {\bibfield  {journal} {\bibinfo  {journal} {Nature
  Methods}\ }\textbf {\bibinfo {volume} {17}},\ \bibinfo {pages} {261}
  (\bibinfo {year} {2020})}\BibitemShut {NoStop}%
\bibitem [{\citenamefont {Ahmed}\ \emph {et~al.}(2021)\citenamefont {Ahmed},
  \citenamefont {Pawar}, \citenamefont {San}, \citenamefont {Rasheed},
  \citenamefont {Iliescu},\ and\ \citenamefont
  {Noack}}]{ahmedClosuresReducedOrder2021}%
  \BibitemOpen
  \bibfield  {author} {\bibinfo {author} {\bibfnamefont {S.~E.}\ \bibnamefont
  {Ahmed}}, \bibinfo {author} {\bibfnamefont {S.}~\bibnamefont {Pawar}},
  \bibinfo {author} {\bibfnamefont {O.}~\bibnamefont {San}}, \bibinfo {author}
  {\bibfnamefont {A.}~\bibnamefont {Rasheed}}, \bibinfo {author} {\bibfnamefont
  {T.}~\bibnamefont {Iliescu}}, \ and\ \bibinfo {author} {\bibfnamefont
  {B.~R.}\ \bibnamefont {Noack}},\ }\href {\doibase 10.1063/5.0061577}
  {\bibfield  {journal} {\bibinfo  {journal} {Physics of Fluids}\ }\textbf
  {\bibinfo {volume} {33}},\ \bibinfo {pages} {091301} (\bibinfo {year}
  {2021})}\BibitemShut {NoStop}%
\end{thebibliography}%

\end{document}